\documentclass[11pt,aps,preprint]{revtex4}
\usepackage{graphicx}
\usepackage{epsfig}
\usepackage{amssymb}
\def\Vec#1{\mbox{\boldmath $#1$}}
\def\beq{\begin{equation}}
\def\eeq{\end{equation}}

\def\beqy{\begin{eqnarray}}
\def\eeqy{\end{eqnarray}}

\newcommand \Pomeron {I\!\!P}
\usepackage{epstopdf}
\usepackage{color}

\begin{document}

\title{Revealing  flickering of the interaction strength 
in $pA$ collisions at the LHC}

\author{M. Alvioli}
\affiliation{Consiglio Nazionale delle Ricerche, Istituto di Ricerca per la 
Protezione Idrogeologica,
  via Madonna Alta 126, I-06128 Perugia, Italy}
  \author{L. Frankfurt}
\affiliation{Tel Aviv University, Tel Aviv, Israel}

\author{V. Guzey}
\affiliation{Petersburg Nuclear Physics Institute, 
National Research Center ``Kurchatov Institute'', Gatchina, 188330, Russia}

\author{M. Strikman}
\affiliation{104 Davey Lab, The Pennsylvania State University,
  University Park, PA 16803, USA}

\begin{abstract}
Using the high-energy color fluctuation formalism 
to include 
inelastic diffractive processes  and taking into account the collision geometry and short-range 
nucleon--nucleon correlations in nuclei, we 
assess
 various manifestations of flickering of the parton wave function of a rapid proton in $pA$ 
interactions at LHC energies in soft QCD processes and in the special soft QCD processes accompanying 
hard processes. We evaluate the 
number of wounded nucleons, $N_{coll}$ --- the number of inelastic collisions of projectile, 
in these processes
and find a nontrivial relation between the hard collision rate and centrality.
 We study the distribution over $N_{coll}$ for a hard trigger selecting configurations  in the nucleon with the strength larger/smaller than the average one and
argue that the pattern observed in the LHC $pA$ measurements by CMS and  
ATLAS for jets carrying a large fraction of the proton momentum, $x_p$,  is consistent with the expectation 
that these configurations interact with the strength which is significantly 
smaller  than the average one --- 
a factor of two smaller for $x_p\sim 0.5$.
We also study the leading twist shadowing and the EMC effects for superdense nuclear matter configurations probed in the events with a larger than average number of wounded nucleons. 
We also 
argue that 
taking into account  
energy--momentum conservation does not change the distribution over $N_{coll}$ but suppresses 
hadron production at central rapidities.
\end{abstract}

\maketitle

\section{Introduction}

Recently a very successful proton--lead run has been performed at the LHC which employed several detectors with 
the acceptance of many units in rapidity. It was observed in~\cite{Atlas} that  interpretation of $pA$ data depends 
significantly on whether one uses as the input model for $pA$ interactions the Glauber model, which does not take into 
account  fluctuations of the interaction strength, or the color fluctuation (CF) approach of~\cite{Heiselberg:1991is,
Alvioli:2013vk} and that it is difficult to describe the data without including such fluctuations. 
The color fluctuation formalism takes into account the space--time evolution characteristic for  the 
interaction of composite states in high energy processes  in QED and QCD. 
In particular, the Lorentz slowing 
down 
of interaction implies that 
an
ultrarelativistic  composite projectile  interacts with a target through 
configurations of partons  whose characteristic lifetime (the coherence length) becomes   large at high energies and whose 
interaction strengths with the target, $\sigma$, may significantly vary.  The fact that the projectile can exist in the frozen 
fluctuations/configurations  of partons  with different interaction  cross sections is called ''flickering'' in our paper.   

Small-size 
parton configurations 
with small $\sigma$
in a meson wave function  
were observed in fixed-target data on pion--nucleus collisions at FNAL  and in  electron--nucleus scattering at 
TJNAF (for a recent review, see \cite{Dutta:2012ii}). Fluctuations to "large" nucleon configurations with the  
larger than average  
$\sigma$  are an unambiguous  consequence of the CF approach~\cite{Heiselberg:1991is}.  
Their contribution allows one  to explain 
the significant large-$N_{coll}$ tail in the distribution over the number of inelastic collisions, 
$N_{coll}$, indicated by  the ATLAS data~\cite{Atlas}.

The aim of this paper is to analyze how fluctuations of the interaction strength, the momentum conservation,  
  the composite structure of hadrons, parton--parton correlations in the parton 
wave function of a fast projectile  hadron and presence of the superdense nuclear matter 
configurations reveal themselves in the structure of final states in $pA$  collisions at the LHC energies.

The paper is organized as follows. In Sect.~\ref{sec:fluct} we explain that the large coherence length for the interaction 
of fast nucleons (which is comparable at the LHC energies to the radius of an atom) results  
in the  necessity to 
take into account 
the significant cross section of diffractive processes in 
proton--nucleon
($pN$) collisions.
For 
proton--deuteron ($pd$) collisions,  
this
leads to the Gribov--Glauber model of  nuclear shadowing for  the   total 
$pd$
cross section~\cite{Gribov}.
Employing completeness over diffractively produced states allows one  to include 
effects of inelastic diffraction  in the interaction of projectile with any target.  This approximation leads to the CF 
approach, which provides a constructive method to  calculate the interaction of projectile   with any number of target nucleons.  
In addition, we explain how to include well understood properties of bound states in QCD  into  the formulae of the 
CF approach. We review the basic formalism and present predictions for the distribution over the 
number of inelastically interacting nucleons, $N_{coll}$.  

 In Sect.~\ref{hardsect} we evaluate  fluctuations of $N_{coll} $ 
   due to CF phenomena 
   in 
   soft QCD processes accompanying  a hard trigger.
The developed formalism explicitly satisfies the QCD factorization theorem for 
hard inclusive 
processes and allows us 
to evaluate the rate of hard processes as a function of the number of wounded nucleons $N_{coll}$ taking accurately into 
account the difference of the impact parameter  geometry 
of hard and soft collisions. Significant deviations from the often assumed  linear 
dependence of  the hard rate on $N_{coll}$ are observed. 

In Sect.~\ref{sec:observ} we  discuss several strategies for observing effects  of 
proton flickering in $pA$ collisions 
with a hard trigger.  In particular, we  argue that such studies would  allow one to determine the correlation between the  
$x$ distribution of partons in the nucleon and the overall interaction strength and, in particular, 
to test the hypothesis that 
the proton size is shrinking with an increase of $x$.  We compare distributions over $N_{coll} $ for triggers corresponding to 
the larger/smaller than average interaction strength.  In particular, we find an enhancement of the jet rate for the peripheral collisions 
in which $x$ of the proton is large enough so that smaller than average configurations in the proton are selected.
We discuss a connection of our results to  the recent 
measurements at the LHC using two large acceptance 
detectors  (ATLAS and CMS), which studied the dependence 
of jet production as a function of the centrality, which was 
 defined via the measurement of the transverse energy distribution in the 
nuclear fragmentation region. We argue that the pattern observed for the forward jet production (along the proton direction)  
matches that for the interaction of the 
proton CF with the strength, which is approximately a factor of two smaller than on average.
 
In Sect.~\ref{evolution} we consider effects of 
perturbative QCD 
(pQCD) evolution on 
color fluctuations for 
fixed-$x$
configurations in the nucleon.
We 
evaluate the range of $x$ at the low $Q$ 
scale 
contributing to
the strength of fluctuations for the same $x$  at the hard probe scale 
 of the order of 100 GeV.

In Sect.~\ref{7}  we consider effects due to 
deviation of nuclear parton distribution functions (PDFs)
from the sum of nucleon PDFs which were neglected in the previous sections 
since they are small in the currently studied kinematics.
We focus on the limit  when a trigger may select collisions where the number of wounded nucleons 
exceeds significantly the average number of nucleons at small impact parameters, 
which---due to the significantly higher local density---corresponds to selection
of configurations in the nucleus wave function for which the parton distribution 
is different from the average one.  
We  demonstrate that for 
these collisions, both nuclear shadowing and the EMC effect are significantly enhanced, 
with the EMC effect probing local nucleon densities  comparable to those  
in the cores of neutron stars.

In the Appendix we explain how to  implement energy--momentum conservation  according to general principles of 
QCD and show that the formulae of the Gribov--Glauber model and the CF  approach   for the total cross section 
and the number  of wound nucleons  are not modified. At the same time,  the formulae for  the double  hadron multiplicity  
in $pd$  collisions (as well as for triple and higher multiplicities in $pA$ scattering)   
are modified by  a model-dependent factor due to an increase of the inclusive cross section with energy. 
This  leads to violation  of the Abramovski--Gribov--Kancheli (AGK) cutting rules~\cite{AGK}  for the inclusive hadron cross section.

\section{Color fluctuations formalism for hadron--nucleus collisions at high energies}
\label{sec:fluct}

In this section we summarize the  framework for the quantitative description of flickering 
phenomena in high energy processes which we refer to as the color fluctuation (CF) approach. This 
framework allows one to take into account  the contribution of the diffractive excitation of a
projectile proton and  implement well-understood  QCD properties of hadrons and their 
interactions. One of such properties is presence of the significant
fluctuations of the interaction strength, 
for a more detailed discussion, see \cite{Frankfurt:2000tya}.   Several types of fluctuations 
are known at present: fluctuations of the sizes and  the shapes of the colliding hadrons, of number 
of interacting constituents, etc.  Following our previous papers  
we will refer to all these fluctuations as color fluctuations.  In the physics of fluctuation 
phenomena, a significant part of fluctuation effects can be evaluated in terms of the dispersion 
of the interaction strengths which is calculable in terms of the cross section of inelastic 
diffractive processes in $pN$ scattering,  see Eq.~(\ref{diffr1}) below.  

It has been understood long ago that in the case of high energy processes, the contribution of the planar Feynman diagrams 
relevant for the Glauber approximation in non-relativistic quantum mechanics is zero and that the dominant contribution 
arises from non-planar diagrams~\cite{Mandelstam,PomeronCalculus}.  Gribov suggested~\cite{Gribov}  to  rewrite the  
sum of non-planar diagrams as the sum over diffractively produced  hadronic intermediate states.   

The Gribov--Glauber model has been further generalized to take into account effects of the
compositeness of a projectile hadron in inelastic  interactions with nuclei~\cite{Heiselberg:1991is}.
This generalization is justified because at high energies it is possible  to neglect effects of 
$t_{min} \neq 0$  ($t_{min}$ is the minimal kinematically allowed four-momentum transfer squared)   
in the production of diffractive excitations of the projectile and,  hence, to  sum over produced diffractive 
states using the condition of completeness:  $\sum_{n} \left|n\right>\left<n\right|=I$, where $I$ is a unit 
matrix. 
The requirement of small  $-t_{min}\le 3 /R_A^2$ puts a limit on the masses of the intermediate states,   
$M_{diff}$, which in the case of nuclei corresponds to
\beq 
M^2_{diff}/s < \sqrt{3}/R_Am_N \,,
\eeq 
or, equivalently, to the configurations in the projectile proton frozen over the coherent length $l_c$:
\beq
l_{coh} ={s\over m_N(M^2_{diff}-m_N^2)} \gg 2R_A \,,
\eeq
where $m_N$ is the nucleon mass; $R_A$ is the nucleus radius; $s$ is the square of the center-of-mass energy.

The first step in the derivation of CF formulae is to notice that  the strength of the interaction with $n$ 
nucleons is modified as compared to the Glauber model  by the factor of 
$\lambda_n \equiv \left<\sigma^n \right>/ \sigma_{tot}^n$  which sums contributions of all diffractive 
intermediate  states.     
The factors of $\lambda_n$ can be expressed in terms of the distribution over   cross sections 
$P_N(\sigma)$,  $ \lambda_n= \int_0^{\infty} d\sigma (\sigma/\sigma_{tot})^n P_N(\sigma)$, where 
$P_N(\sigma)$ is  the probability for a proton  to interact with the target with the given cross section $\sigma$
and $\sigma_{tot}=\int_0^{\infty} d\sigma \sigma P_N(\sigma)$ is the total proton--nucleon cross section.
The distribution  $P_N(\sigma)$ depends on the incident energy, which will be discussed later.
 
By construction,  $\lambda_0=\lambda_1=1$ due to the probability conservation and the definition of the total cross 
section. The variance of the distribution $P_N(\sigma)$ is 
\begin{equation}
\lambda_2 - 1= \int_{0}^{\infty} d\sigma \, P_N(\sigma) \left({\sigma\over \sigma_{tot}} -1\right)^2\equiv \omega_{\sigma}= 
\left.{{d\sigma(p+ p\to X+p) \over dt}\over { d\sigma(p+ p\to p+p) \over dt}}\right\vert _{t=0},
\label{diffr1}
\end{equation}
where  the sum over diffractively produced states $X$ is implied.   Equation~(\ref{diffr1}) follows directly from the 
optical theorem and the definition of $P_N(\sigma)$. It was derived originally in  \cite{Miettinen:1978jb} within the 
approach of \cite{Good:1960ba}.  The analysis of the  fixed target data \cite{Blaettel:1993rd}  indicates that the 
variance $\omega_{\sigma}$ first grows with energy reaching $\omega_{\sigma} \sim 0.3$ for $\sqrt{s} \sim$ 100 GeV 
and then   starts 
to decrease at higher energies dropping to $\omega_{\sigma} \sim 0.1 $  at the LHC energies.

Thus in contrast to the case of lower energies, the
cross section is  calculable in terms of scattering of frozen parton 
configurations in the wave function of a rapid projectile and then summing over contributions of these configurations. 
In the case of averaging of quantities depending on one variable $\sigma$, we may introduce the following unit matrix,
$\int d\sigma \delta(\sigma-\sigma(x_i,\rho_{i,t}))$, where $x_i$ and $\rho_{i,t}$ are the light-cone fractions and transverse 
coordinates of the partons, integrate  over all variables characterizing the wave function of the projectile, $\psi$, and obtain:
\begin{equation}
\int |\psi(x_i,\rho_{i,t})|^2\delta(\sigma-\sigma(x_i,\rho_{i,t})) d\tau=P_N(\sigma) \,,
\label{fluent}
\end{equation}
where $d\tau$ is the phase volume. This formula indicates that selection of certain parton configuration in the projectile may 
influence the effective value of $\sigma$.  Note also that the contribution of large diffractive masses described by triple Pomeron 
processes is restricted by the kinematics $M_{diff}^2/s \ll 1/m_N R_A$. Thus, at large $s$,  participating 
parton configurations within the projectile are frozen during collisions   and the contribution of large diffractive masses  can 
be included in Eq.~(\ref{fluent}).    This means that the CF approach  also includes 
large diffractive masses corresponding to  triple Pomeron processes.

Important properties of $P_N(\sigma)$  follow from rather general reasoning: \\
\noindent
(i)\, 
$P_N(\sigma)$  is positive and rapidly  decreasing with an increase of $\sigma$  to ensure finiteness of the moments 
$\int P_N(\sigma) \sigma^{n} d\sigma $.   

\noindent
(ii)\,  $P_N(\sigma)$  is a continuous function of $\sigma$ with  $P_N(0)=P_N(\sigma\to \infty) =0$,
which follows from applicability of pQCD at $\sigma\to 0$,   see the discussion below.
 Hence, $P_N(\sigma)$  should have a maximum at $\sigma=\sigma_0$ 
 corresponding to an average configuration of partons  
 in the nucleon,  see Eq.~(\ref{psigma}) below.  
 Thus,  $\sigma_0$ is close to the observed total nucleon--nucleon ($NN$) cross section.

\noindent
(iii)\,
The distribution over $\sigma$ around  the average configuration is controlled by  the variance $\omega_{\sigma}$.
The variance is expressed in terms of  the cross section of inelastic diffraction at $t=0$, see Eq.~(\ref{diffr1}).

The data indicate that the variance first grows with energy reaching $\omega_{\sigma} \sim 0.3$ 
for $\sqrt{s} \sim$ 100 GeV and then starts to decrease for higher energies. The current LHC data on diffractive 
processes in $pp$ collisions are not sufficient to determine accurately  $\omega_{\sigma}$ directly from 
the data. Still the data are consistent with the trend that the interaction at small impact 
parameters becomes practically black and hence does not lead to inelastic  diffraction. Overall, 
extrapolations from  the lower energies and an inspection of preliminary LHC data  indicate
that  the ratio of diffractive and elastic cross sections at $t=0$ drops with energy and that 
$\omega_{\sigma}(\sqrt{s}= \mbox{5 TeV}) \approx 0.1$. (This is close to the 
extrapolation of the pre-LHC data fit for $\omega_{\sigma}$ by  K.~Goulianos \cite{Goulianos} to LHC energies.)
Naively this looks like a small number but it corresponds to a rather broad distribution 
over $\sigma$. For example,   modeling $P_N(\sigma)$ by introducing two diffractive states of equal probability, 
one would find that they
should have the cross sections that differ by nearly a factor of two:
$\sigma_i=\sigma_{tot}  (1\pm  \sqrt{\omega_{\sigma}})$.  This indicates that even at the LHC, the nucleon 
can interact with a significant probability both with the super large strength $\sim$ 130 mb and  the significantly 
smaller than average strength  $\sim$ 70 mb.

\noindent
(iv)\,
In the region of  large $\sigma$ one can use 
several 
generic considerations.   Since the variance  is small at the LHC energies, the distribution around the maximum is comparatively 
narrow and in  practical calculations, the region of $\sigma\gg \sigma_0$ gives a negligible contribution. Thus, a reasonable  
approximation is take into account only small fluctuations around the average value of $\sigma$.   
Then, as in the classical and quantum mechanical theory of  small fluctuations around average value,
$P_N(\sigma)$ in the vicinity of $\sigma=\sigma_0$  should have the form close to the Gaussian distribution in $\sigma$.
   Note also that models with different  patterns of fluctuations    
such as, e.g., the model with two cross section eigenstates  but the same 
$\omega_{\sigma}$ \cite{Alvioli:2013vk} and the model with
$P_N\propto \exp[-c\left|\sigma-\sigma_0\right|/\sigma_0] $   lead to very similar numerical results.

\noindent
(v)\,
At small $\sigma$, $P_N(\sigma)\propto \sigma$ which  follows from  QCD quark models of the proton  and approximate 
proportionality of the cross section of interaction of small-size   $\left|3q\right>$ configurations with target nucleons to 
the area occupied by color as follows from pQCD, see e.g.\cite{Low}.  Under these assumptions, the derivation is effectively 
reduced to the application of QCD quark counting rules.   Note that for a projectile meson, $P_{\pi}(\sigma) \propto const$ at 
$\sigma\to 0$  \cite{Blaettel:1993rd}.    In perturbative QCD, the interaction cross section of small-size configurations 
is small but grows with energy faster than that of average-size configurations. 
As a result, $P_N(\sigma)$ is expected  to decrease rather rapidly with $s$  for fixed $\sigma \ll \sigma_0$.

A competing parametrization of $P_N(\sigma)$ based on the Poisson distribution 
has been suggested in \cite{Coleman-Smith:2013rla} for RHIC 
energies. In this parametrization, $P_N(\sigma)\propto \sigma^{k-1}\exp(-\sigma/\theta) $, with 
$k= 1/\omega_{\sigma}$. For RHIC (LHC), where $\omega_{\sigma}\approx 0.25$ $(0.1)$,  this corresponds to 
$P_N(\sigma)_{|\sigma\to 0}\propto \sigma^3 (\sigma^9)$, which is much faster than in  the quark models 
where $P_N(\sigma)_{|\sigma\to 0}\propto \sigma$.

\noindent
(vi)\,
 The resulting form  of $P_N(\sigma)$ is a smooth interpolation between 
 the small-$\sigma$ and large-$\sigma$  regimes. In our 
 numerical studies  we  use results of 
 the theoretical analysis of \cite{Blaettel:1993rd}  which determined 
first three moments of $I_n=\int \sigma^n P_N(\sigma) d\sigma$ using the normalization condition for 
$P_N(\sigma)$, Eq.~(\ref{diffr1}) for the variance and the  data on coherent diffraction off the deuteron  and 
implemented the small-$\sigma$ behavior of $P_N(\sigma)$ expected in pQCD:
\beq
P_N(\sigma)\,=\,\gamma\,\frac{\sigma}{\sigma+\sigma_0}
\exp\left\{-\frac{(\sigma/\sigma_0\,-\,1)^2}{\Omega^2}\right\},\,
\label{psigma}
\eeq 
where $\Omega^2/2\approx \omega_{\sigma}$ numerically. In the $\omega_{\sigma}\to 0$ limit, $\Omega^2= 2 \omega_{\sigma}$ 
and the parametrization of Eq.~(\ref{psigma}) converges to $\delta(\sigma -\sigma_{tot})$.   The analysis \cite{Blaettel:1993rd}  
of the data on coherent diffraction off the deuteron at $E_p \mbox{= 400 GeV} $  shows that this distribution is 
approximately symmetric around $\sigma = \sigma_{tot}$.

Equation~(\ref{psigma})  is qualitatively  different  from $P_N(\sigma)$ suggested in 
\cite{Miettinen:1978jb} to  describe $pN$ scattering using
the pre-QCD idea that only wee partons are involved in the high energy hadron--hadron interaction. 
In particular, instead of the behavior $P(\sigma\to 0) \propto \sigma$, the authors of~\cite{Miettinen:1978jb} 
suggested  that $P_N(\sigma\to 0 ) \propto \delta(\sigma)$.

 For $pA$ collisions at $\sqrt{s}$=5.02 TeV   studied at the LHC, we use $\sigma_{tot}=93$ mb and $\omega_{\sigma}=0.1$ 
leading to  $\gamma=0.0263914$, $\sigma_0=86.4825$~mb, and $\Omega=0.51285$.
Although experimentally the value of $\omega_{\sigma} =0.1$ appears to be 
preferred for the energies probed in $pA$ collisions at the LHC, in 
several cases we will illustrate sensitivity to the value of $\omega_{\sigma}$ by presenting numerical 
results also for $\omega_{\sigma} =0.2$.

To evaluate the cross sections of the events where the number of collisions is exactly $N_{coll}$, 
where $N_{coll}$ is the number of  target nucleons involved in the inelastic   interaction  with the projectile,
one needs the distribution over inelastic cross sections,  $P_{inel}(\sigma_{inel})$. This distribution is 
calculable in terms of $P_N(\sigma)$ since all above discussed restrictions on $P_{inel}(\sigma_{inel})$ and 
$P_N(\sigma)$  are  the same except for the normalization. Moreover, since experimentally  the fraction 
$1-\lambda$ of the total cross section due to elastic scattering is a rather weak function of the incident energy, 
it is natural to  assume that this is also true for each individual proton fluctuation.
Thus, the variances of  $P_{inel}(\sigma_{inel})$ and $P_N(\sigma)$ are equal and one can restore 
$P_{inel}(\sigma_{inel})$ using the following relation:
\begin{equation}
P_{inel}(\sigma_{inel}=\sigma \frac{\sigma_{inel}}{\sigma_{tot}} )=\frac{\sigma_{inel}}{\sigma_{tot}} P_N(\sigma) \,.
\label{eq:P_sigma}
\end{equation}

In the discussed approximation one can rewrite $\sigma_{in}(pA)$  as  a
sum of positive cross sections of inelastic interactions with exactly $N_{coll}$ nucleons analogously to the case of 
the Gribov--Glauber  approximation of \cite{Bertocchi:1976bq}. A compact expression for $\sigma_{in}(pA)$
 can be written, if internucleon 
correlations in the nucleus and the finite radius  of the $NN$ interaction are neglected:
\begin{eqnarray} 
  && \sigma^{hA}_{in} = \sum^{A}_{N_{coll} = 1} \sigma_{N_{coll}} \,, \nonumber\\
  && \sigma_{N_{coll}} =\int d\sigma P_{inel}(\sigma)  \frac{A!}{(A - N_{coll})!\, N_{coll}!} \int
  d^2\Vec{b}\, x(b)^{N_{coll}} \left[1-x(b)\right]^{A-N_{coll}} \,,
  \label{eq3}
\end{eqnarray} 
where $x(b)=\sigma T(b)/A$ and the normalization is $\int d^2 \Vec{b}\,T(b)=A$. This formula is a generalization of the 
optical approximation to the relativistic domain where inelastic processes give the dominant contribution to the total cross 
section. In the framework of the Gribov Reggeon calculus, the factor of 
$x(b)^{N_{coll}}$ corresponds to $N_{coll}$ cut  Pomeron exchanges and 
the factor of $ \left[1-x(b)\right]^{A-N_{coll}}$ --- to $A- N_{coll}$  uncut  Pomeron exchanges. 

It is straightforward to include the effect of the  finite radius of the $NN$ interaction as the probability for two nucleons  
to interact inelastically while at a relative impact parameter $b_{12}$,  $P(b_{12})$. It  is expressed through the profile 
function of $NN$ scattering, $\Gamma(b_{12})$,  as $P(b_{12})= 1 -\left|1- \Gamma(b_{12})\right|^2$. The resulting formula 
(an analogue of Eq.~(25) of \cite{Bertocchi:1976bq} written in the approximation when correlations between nucleons are 
neglected)  is essentially probabilistic reflecting the semiclassical picture of 
high energy inelastic interactions with nuclei. The Monte Carlo (MC) which includes 
accurately both geometry of the $NN$ interactions and nuclear correlations 
was presented in \cite{Alvioli:2013vk}. It is used  in our numerical studies  described below.

Hence the probability of inelastic collisions with exactly $N_{coll}$ nucleons, $P_{N_{coll}}$,  is simply:
 \begin{equation} 
P_{N_{coll}}=  \sigma_{N_{coll}}/\sigma^{hA}_{in} \,.
\end{equation} 
For the average number of collisions, one finds 
\beq
\left<N_{coll}\right> = \sum^{A}_{N_{coll} = 1}N_{coll}\sigma_{N_{coll}}/\sigma^{hA}_{in}=A\sigma_{in}/\sigma^{hA}_{in} \,,
\label{Naver}
\eeq
which depends very weakly  on $\omega_{\sigma}$ \cite{Alvioli:2013vk} since the inelastic shadowing correction to 
$\sigma^{hA}_{in}$ is very small because the $pA$ interaction is nearly black  at the LHC energies.  

In terms  of non-planar diagrams, energy--momentum conservation is automatically fulfilled. This implies  that 
taking into account  energy--momentum conservation does not produce additional factors in the formulae of 
the Gribov--Glauber and CF approaches  for  the total cross sections, inelastic shadowing and hadron multiplicities  
at the rapidities close to the nucleus fragmentation regions.  In contrast, the formulae for the double, triple, etc.~hadron 
multiplicities contain   additional suppression factors  to satisfy energy--momentum conservation, see the discussion in the Appendix.

In the approximation  of \cite{Bertocchi:1976bq},   $\sigma_{inel}$ did not include inelastic final states 
with the nucleus breakup but without  hadron production. Correspondingly, in our case, when a particular configuration 
can scatter elastically off a nucleon of the nucleus,  the final states corresponding to the excitation of the projectile 
without hadron production of the nucleus fragmentation are not included in $\sigma_{inel}$ in Eq.~(\ref{eq3}) (or its finite 
radius of interaction version).  Namely,  Eq.~(\ref{eq3})
does not include the cross section of coherent inelastic diffraction, which is  less than 1\% 
of the total inelastic cross section \cite{Guzey:2005tk}, and quasielastic scattering with the nucleus breakup.
Incoherent diffraction  is dominated by scattering off the nucleus edge which is roughy 
equal to the product of the probability of the interaction  with one nucleon ($\sim 20\%$, see Fig.~1) 
and the probability of single diffraction for  a given proton in inelastic $pp$ collisions,
which  is $\sim 10-15$\%, leading to the overall probability of incoherent diffraction of $\sim 2 - 3\%$.

These contributions are also  not included in the LHC $pPb$ events samples  -- events without rapidity gaps.
 This allows one to exclude the Coulomb excitation contribution 
which may reach 10\% of the inelastic cross section  \cite{Guzey:2005tk}. This cut removes from the sample also 
most of the  rapidity gap events  due to the inelastic diffraction dissociation of the proton and the nucleus.
A type of the events which is  included in our definition of $\sigma_{inel}$ but not in the experimental definition is 
quasielastic scattering in which nucleon (nucleons) of the nucleus are diffractively excited.   In principle one needs to 
include this correction in the comparison of the calculations with the data, although as we have seen above, in 
most of the cases it is a very small effect.

The finite radius of the nucleon--nucleon interactions  and short-range $NN$ correlation effects were 
implemented in the  Monte Carlo procedure of Ref.~\cite{Alvioli:2013vk}. The algorithm generates multi-nucleon 
configurations in nuclei with correct short-range correlations of protons and neutrons developed in~\cite{Alvioli:2009ab} and
uses the profile function for the dependence of the probability of inelastic 
$NN$ collisions on the relative impact parameter as given by the Fourier transform of the elastic $pp$ 
amplitude and $S$-channel unitarity. Fluctuations of the interaction strength are 
included by assigning incoming  protons the values of $\sigma$ with the measure 
given by $P_N(\sigma)$. 

\begin{figure}[htp]
  \vskip -0.0cm
  \centerline{\includegraphics[width=10.0cm]{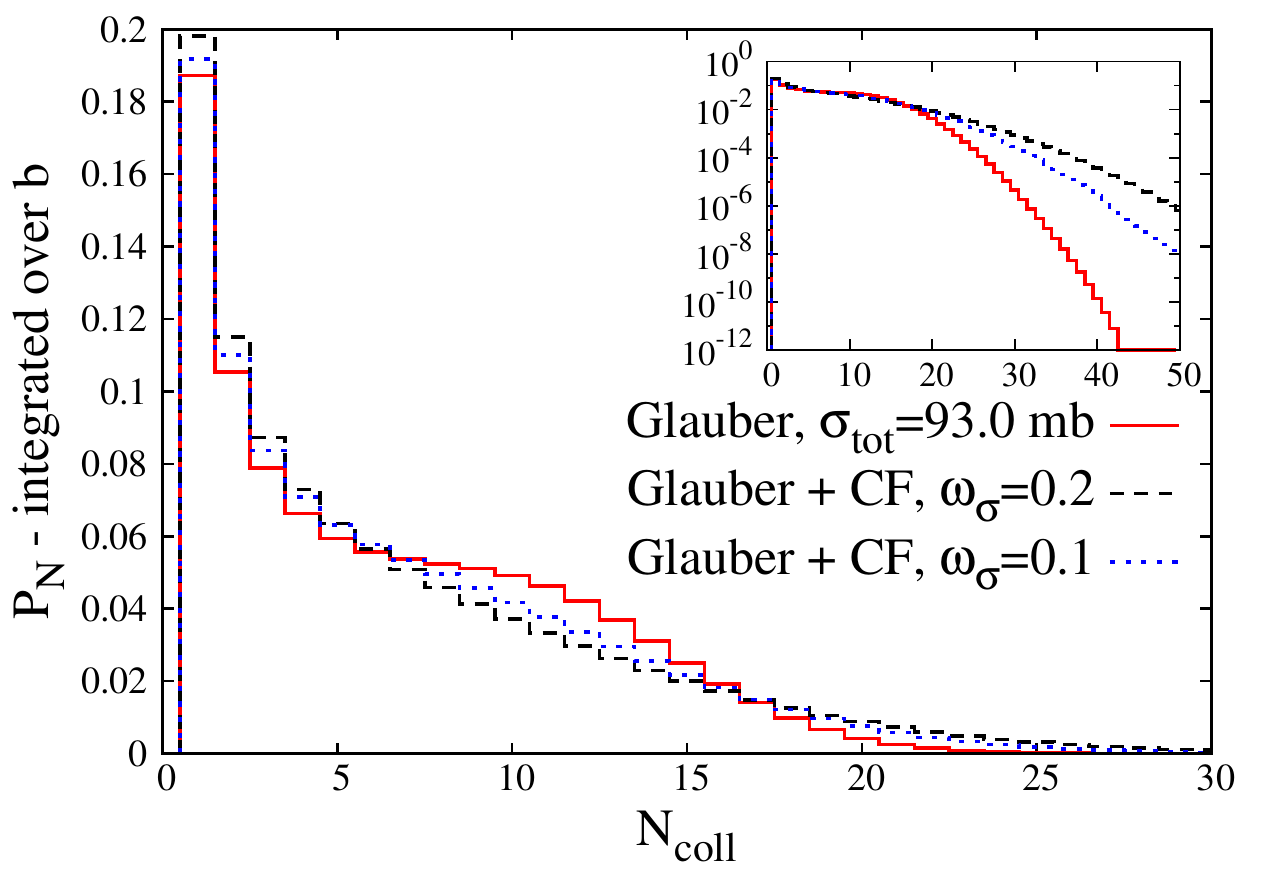}}
  \caption{The probabilities $P_N$ of having $N=N_{coll}$ wounded nucleons, 
averaged over the global impact parameter $b$, as a function of $N_{coll}$ for the Glauber 
model ($\omega_{\sigma}=0$) and in the CF model with  $\omega_{\sigma}=0.1$ 
(our base value used in the current analysis)  and $\omega_{\sigma}=0.2$.
The inset is in the log scale.}
  \label{fig1}
\end{figure}

In Ref.~\cite{Alvioli:2013vk}  a detailed comparison of the predictions for the number of wounded 
nucleons with and without taking into account color fluctuations was presented. 
It was demonstrated  that the inclusion
of fluctuations leads to a significant change of the distribution over the number 
of wounded nucleons both for a fixed impact parameter and for the integral over impact parameters.
A large enhancement of the probability of the events with large $N_{coll}$ was observed 
(see Fig.~\ref{fig1}).

As usual for the random  phenomena,  in a wide range of $N_{coll}$, the probability distribution over 
$N_{coll}$ ($P(N_{coll})$) is most sensitive to the value of the variance $\omega_{\sigma}$. 
In particular, the parametrization of Eq.~(\ref{psigma}) and the two-state model were found to give 
very close results in a wide range of $N_{coll}$.

The results of our numerical studies using the Glauber model 
(corresponding to $\omega_{\sigma}=0$)  and the CF 
model with two values of $\omega_{\sigma}$
($\omega_{\sigma}=0.1$ and 0.2)
 are presented in Fig.~\ref{fig1}. 
The calculation is done using the  Monte Carlo algorithm developed by two of the present authors  
and described previously \cite{Alvioli:2013vk}. 
The profile function was also scaled  with $\sigma$ to satisfy the condition that the interaction 
is black at small impact parameters. 
One can see from the inset of Fig.~\ref{fig1} that our analysis
demonstrates that the distribution over $N_{coll}$ is  sensitive 
to the value of  $\omega_{\sigma}$ and that fluctuations result in the substantially 
larger tail of the distribution at large $N_{coll}$.
 
\begin{figure}[htp]
  \vskip -0.0cm
  \centerline{\includegraphics[width=10.0cm]{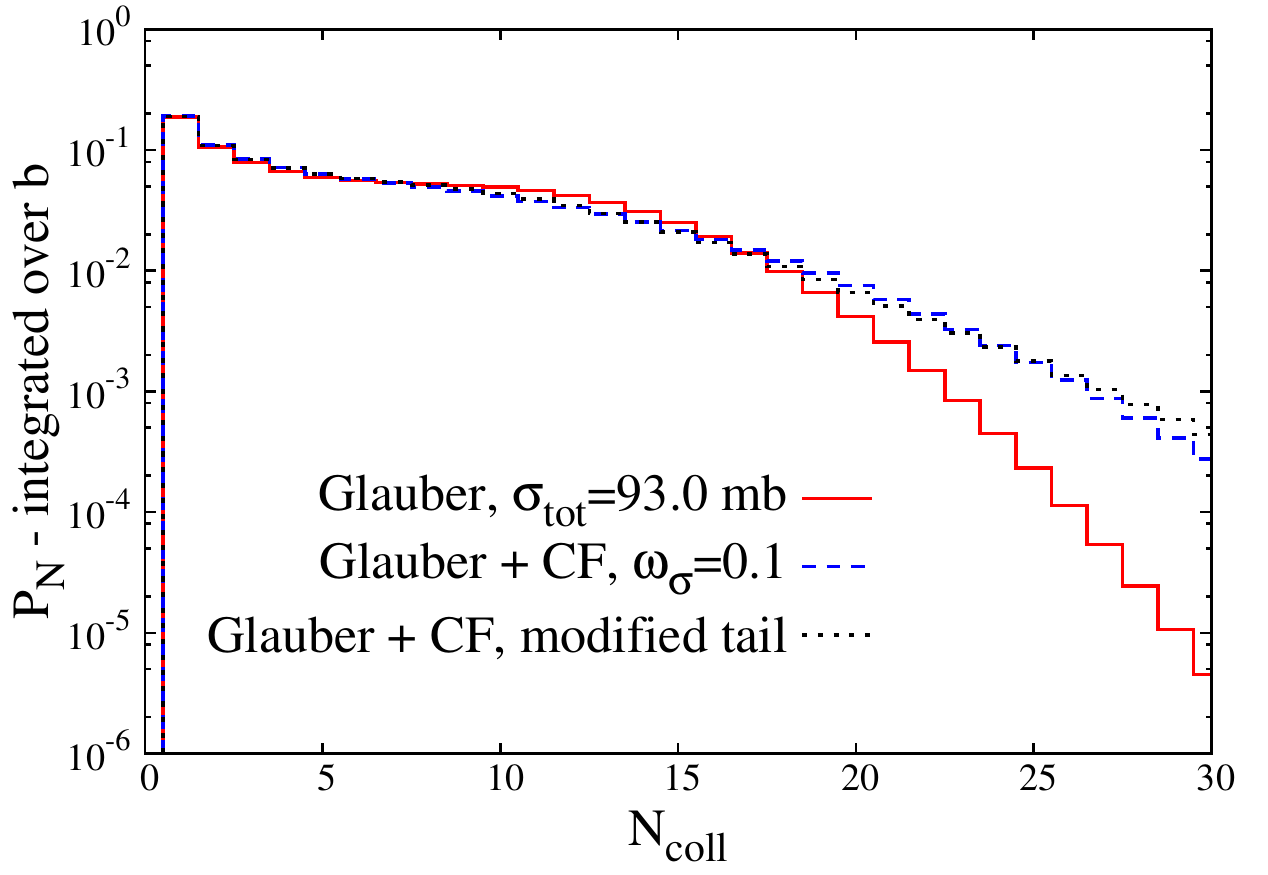}}
  \caption{Comparison of the distributions over $N=N_{coll}$ for the Glauber model and for
  the color fluctuation model with $\omega_{\sigma}= 0.1$
   with
  the Gaussian [Eq.~(\ref{psigma})] and exponential [Eq.~(\ref{linexp})] large-$\sigma$ behavior.}
  \label{lingauss}
\end{figure}

In Fig.~\ref{fig1} we showed the results of calculations 
based on the parametrization  suggested  in~\cite{Blaettel:1993rd}, which  
assumes the Gaussian shape of the large-$\sigma$ tail of $P_N(\sigma)$.   However, since the study~\cite{Blaettel:1993rd} 
was testing fluctuations near its average value, $\sigma_{tot}$,   it is reasonable to consider other options for large-$\sigma$ asymptotic  of  $P_N(\sigma)$ in the present work.  In particular,
the tail of small-$x$ parton distributions in the transverse plane is often fitted by the Gaussian distribution in $\rho^2$,
where $\rho$ is the parton transverse coordinate.   If 
the cross section for large $\rho$   is approximately proportional to the area, i.e.,  $\sigma \propto \pi \rho^2$, one would 
expect presence of the large-$\sigma$ tail of $P(\sigma)$ 
that behaves as   $P(\sigma ) \propto \exp(-c \sigma)$. To probe sensitivity to the possible
 presence of such a tail, we introduce another model of  $P_N(\sigma)$:
\begin{equation}
  P_N(\sigma)= a \sigma \exp(-c \left|\sigma -\sigma_0\right|) \,,
  \label{linexp}
\end{equation}
with parameters fixed to reproduce the same total cross section and dispersion as in the basic model.  
We find that the distribution over $N_{coll}$ practically does not change -- see Fig.~\ref{lingauss}.

This confirms the conclusion of~\cite{Alvioli:2013vk} based on the comparison of the model based 
on Eq.~(\ref{psigma}) and the two-component model.  
At the same time, changing the behavior at small 
$\sigma$ one can generate a very different shape for the same variance,
see Ref.~\cite{Coleman-Smith:2013rla}.  Hence  it would be 
interesting to explore this issue further as the sensitivity to the tail for central collisions 
should grow since at the LHC in central $pA$ collisions,  one typically selects  $N_{coll} \sim 14$.

\section{Distribution over the number of collisions for processes with a hard trigger}
\label{hardsect}

We begin by addressing 
the long-standing question of the interplay of the phenomenon of color 
fluctuations and the partonic  structure of the nucleon. It is well understood and observed 
experimentally that a hadron can exist in the configurations of  different transverse 
sizes and that smaller configurations interact with a smaller cross section than the larger  
size configurations.  This is one of the origins of flickering  of the interaction strengths, 
which, as we mentioned in the Introduction, is present in both QCD and QED.  Note here that a 
related phenomenon of fluctuations of the nucleon gluon density at fixed small $x$  was 
inferred from exclusive hard processes   in~\cite{Frankfurt:2008vi}.
One of the typical setups for $pA$ collisions is the study of soft phenomena which accompany 
a hard subprocess (dijet, $Z$-boson, $\dots$)  and is related to the number of wounded nucleons.

Our main aim is to get a deeper insight into dynamics of 
$pA$ interactions   and in particular  to probe the flickering phenomenon which we discussed in the Introduction.
In the case of inclusive production, the cross section is given by the  QCD  factorization theorem. 
An additional requirement on the final state breaks down the closure  approximation and hence 
requires another  form of the factorization theorem.

In this section we will consider nuclear PDFs as a sum of the nucleon PDFs since nuclear effects are small for large $p_t$ studied at the LHC except 
possibly in the region of $x_A \ge 0.4 $ where the EMC effect may play a role.
Correspondingly we  will use the  impulse approximation to evaluate cross sections of hard process and the
CF approach to calculate the number  of wound nucleons  accompanying the hard process. Effects related to the deviations 
of the nuclear PDFs from the additive sum of the nucleon PDFs---leading twist nuclear shadowing and the EMC effect---will be considered in Sect.~\ref{7}.

On average, in the geometric model for hard processes in the kinematics, where nuclear shadowing can be 
neglected, i.e.,  for $x\ge 0.01$  and  even smaller $x$ for large virtualities, 
the multiplicity of events with a hard trigger (HT),  which we will denote as $ Mult_{pA} (HT)$, is 
$ Mult_{pA} (HT)=\sigma_{pA}(HT +X)/\sigma_{pA}(in)= A\sigma_{pN}(HT +X)/\sigma_{pA}(in)$. Using 
$ Mult_{pN} (HT)=\sigma_{pN}(HT +X)/\sigma_{pN}(in)$
and 
Eq.~(\ref{Naver}) (which to a very good approximation holds in the CF approximation \cite{Alvioli:2013vk} ) one finds that 
a simple relation for the multiplicities of HT events in $pN$ and minimal bias $pA$ collisions holds:
\begin{equation}
  Mult_{pA} (HT) =  \left<N_{coll}\right> Mult_{pN} (HT) \,.
  \label{multav}
\end{equation}

Here we will consider the rates of hard collisions  as a function of $N_{coll}$ with the additional factor of
$N_{coll} $ in the denominator in order to focus on the deviation from the naive optical model expectation 
\cite{Frankfurt:1985cv}  that Eq.~(\ref{multav}) holds for fixed values of $N_{coll}$:
\begin{equation}
R_{HT}(N_{coll}) \equiv  {Mult_{pA} (HT)  \over Mult_{pN} (HT) N_{coll}} = 1 \,.
  \label{rnm}
\end{equation}

The impact parameter  dependence of the 
cross section for the hard collision of two hadrons follows from QCD factorization theorem. It is given 
by the convolution of two generalized parton distributions which are functions of $\Vec{\rho}_1$ and $\Vec{\rho}_2$ -- transverse distances of partons from the center of mass of the corresponding hadrons -- with condition $\Vec{\rho}_1+\Vec{b} -\Vec{\rho}_2=0$ with accuracy $1/ p_t(jet)$.  When further integrating over $\Vec{b }, \Vec{\rho}_1,\Vec{\rho}_2$ one obtains usual collinear expression for the cross section through  the product of the  pdfs of the hadrons, see e.g. discussion in \cite{Frankfurt:2003td}.

\begin{figure}[htp]
  \vskip -8.0cm
  \centerline{\includegraphics[width=10.0cm]{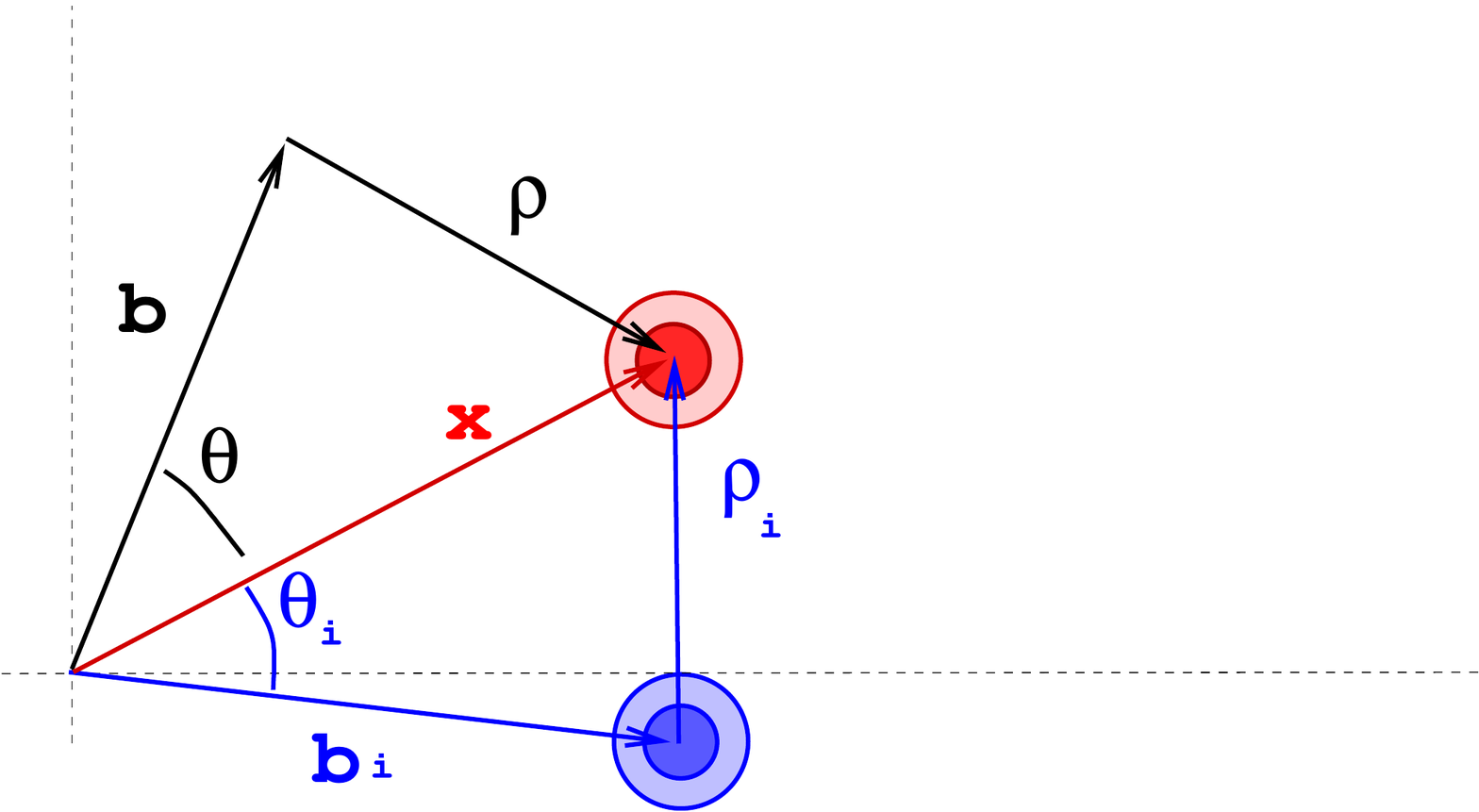}}
  \caption{Sketch of the transverse geometry of collisions.}
  \label{figsketch}
\end{figure}

To describe geometry of dijet production in  proton -- nucleus collisions 
let us  introduce vectors $\Vec{b}$ and $\Vec{b_j}$  the transverse center of mass  of the projectile proton 
and the target nucleons relative to the center of the nucleus, respectively.  We also denote as
$\Vec{\rho}$ the transverse distance of the parton of the projectile from point $\Vec{b}$.
The transverse distance between the point of the hard collision and the distance to the transverse c.m. of 
nucleon $j$ of  the nucleus is
\begin{equation}
\Vec{\rho}_j = \Vec{b} + \Vec{\rho} - \Vec{b}_j \,.
\label{eq1}
\end{equation}
The discussed geometry of collisions is shown in Fig.~\ref{figsketch}.

The generalized gluon  distribution in the nucleon can be parametrized as 
$g_N(x,Q^2, \rho) = g_N(x,Q^2) F_g(\rho)$, where $F_g(\rho) $ is the normalized distribution of gluons 
in the nucleon transverse plane (we do not write here explicitly the dependence of $F_g(\rho)$
 on $x$ and $Q^2$);   $\int d^2\rho F_g(\rho)=1$.
This parametrization is reasonable since the distribution over $\rho$ is practically independent on $Q^2$.    
In our numerical calculations, we take $F_g(\rho)$ from the analysis  of the  data on elastic photoproduction 
of J/$\psi$ mesons  \cite{Frankfurt:2002ka,Frankfurt:2003td,Frankfurt:2010ea}.  For $x\sim 0.01$:
\begin{equation}
 F_g(b)= (\pi B^2)^{-1}\exp\left[-b^2/B^2\right],
\end{equation}
where $B= 0.5$ fm. Note that sensitivity to the exact value of $B$ is rather insignificant as long as $x$ stays
small enough.

The cross section differential in the  impact parameter is given by convolution of the generalized 
gluon distributions of the colliding particles:
\beq
{d\sigma_{HT}(NA) \over d^2b}
=\sigma_{HT}(NN)  \int d^2 \rho\prod_{j=1}^{j=A} [d^2\rho_j] F_g(\rho) \times \sum_{j=1}^{j=A} F_g(\rho_j) \,,
\label{impact}
\eeq
where  $\Vec{\rho}_i$ is given by Eq.~(\ref{eq1}). The averaging over configurations in the nucleus is 
implied but not written explicitly. 

It is worth emphasizing that 
Eq.~(\ref{impact}) automatically  corresponds  to the impulse approximation for the total inclusive cross section of the HT process:
\begin{equation}
  \int d^2b {d\sigma_{HT}(NA) \over d^2b}= A\sigma_{HT}(NN) \,.
\end{equation}
 
Up to this point, the integral over $d^2 \rho$ can be performed analytically (or numerically) since the
integrand function $F_g(\rho)\sum^A_{j=1}F_g({\bf b}+\mathbf{\rho}-{\bf b}_j)$, for a given configuration 
and given $b$, depends on $\rho$ which has to take every possible value inside the nucleus.

However, the calculation of the distribution over $N_{coll}$ involves taking into account that 
much smaller impact parameters dominate in hard collisions than in soft collisions 
\cite{Frankfurt:2003td,Frankfurt:2010ea}. Also we want to  be able to take into account 
correlations of nucleons in nuclei. Consequently the calculation can be performed only using 
a Monte Carlo technique.

The algorithm which leads to the impulse approximation expression for the cross section summed over the contributions of all $N_{coll}$ is as follows.

(i) First  a configuration of nucleons in the nucleus is generated and a particular value of $b$ is chosen.

(ii) The quantity $F_g(\rho) \times \sum_{j=1}^{j=A} F_g(\rho_i) $ gives the weight of these 
configurations to  the average when we calculate the integral over $b$.

(iii) The nucleon involved in the hard interaction is assigned to nucleon $j$ with the probability  
given by 
\begin{equation}
  p_j=\frac{F_g({\bf b}+{\bf\rho}-{\bf b}_j)}{\sum^A_{k=1}F_g(({\bf b}+\mathbf{\rho}-{\bf b}_k))} \,.
\end{equation}

(iv)  The number of other nucleons which interacted inelastically  is calculated  
(that is, all nucleons except nucleon  $j$); this number is  $N_{coll}(other)$. 
This component of the procedure is identical to the one described for a generic calculation 
of $N_{coll}$ without a trigger described in Sect.~\ref{sec:fluct}.  
As a result, we can calculate now 
the probability  that the 
interaction with the generated configuration will lead to  $N_{coll}$ active nucleons:    
\begin{equation}
  N_{coll}= N_{coll}(other) + 1 \,,
\end{equation}
and, hence, determine the probability that in the event there are exactly $N_{coll}$. We denote this probability 
as $p_{hard}(N_{coll},event)$.

(v) Finally we  calculate the rate of the hard collisions  due to events with a specific number of collisions 
(we suppress here the overall factor of $\sigma_{pN}(HT))$:
\begin{equation}
  \label{mark1}
  \int d^2b d^2 \rho\prod_{j=1}^{j=A} [d^2\rho_j] F_g(\rho) \times \sum_{j=1}^{j=A} F_g(\rho_i)p_{hard}(N_{coll},event) \,.
\end{equation}

The fraction of such events is simply
\begin{equation}
  Frac(N_{coll}) = {1\over A}  \int d^2b d^2 \rho\prod_{j=1}^{j=A} [d^2\rho_j] F_g(\rho) 
  \times \sum_{j=1}^{j=A} F_g(\rho_i)p_{hard}(N_{coll},event) \,.
  \label{frac}
\end{equation}

As we explained above, in order to compare with the naive expectation of the Glauber model 
without correlations of any kind and the optical model limit, where one expects that the cross section 
of hard collisions for events with $N_{coll}$ is $N_{coll}\sigma_{hard}(NN)$, we calculated the 
ratio given by Eq.~(\ref{rnm}). This procedure is obviously consistent with 
\begin{equation}
  \sum_{N_{coll}}\sigma(N_{coll})N_{coll}= A\sigma_{NN} \,.
\end{equation}
Note here that in this discussion, we did not address the potential effect of 
energy--momentum conservation, see the Appendix.

\begin{figure}[htp]
  \vskip -0.0cm
  \centerline{\includegraphics[width=10.0cm]{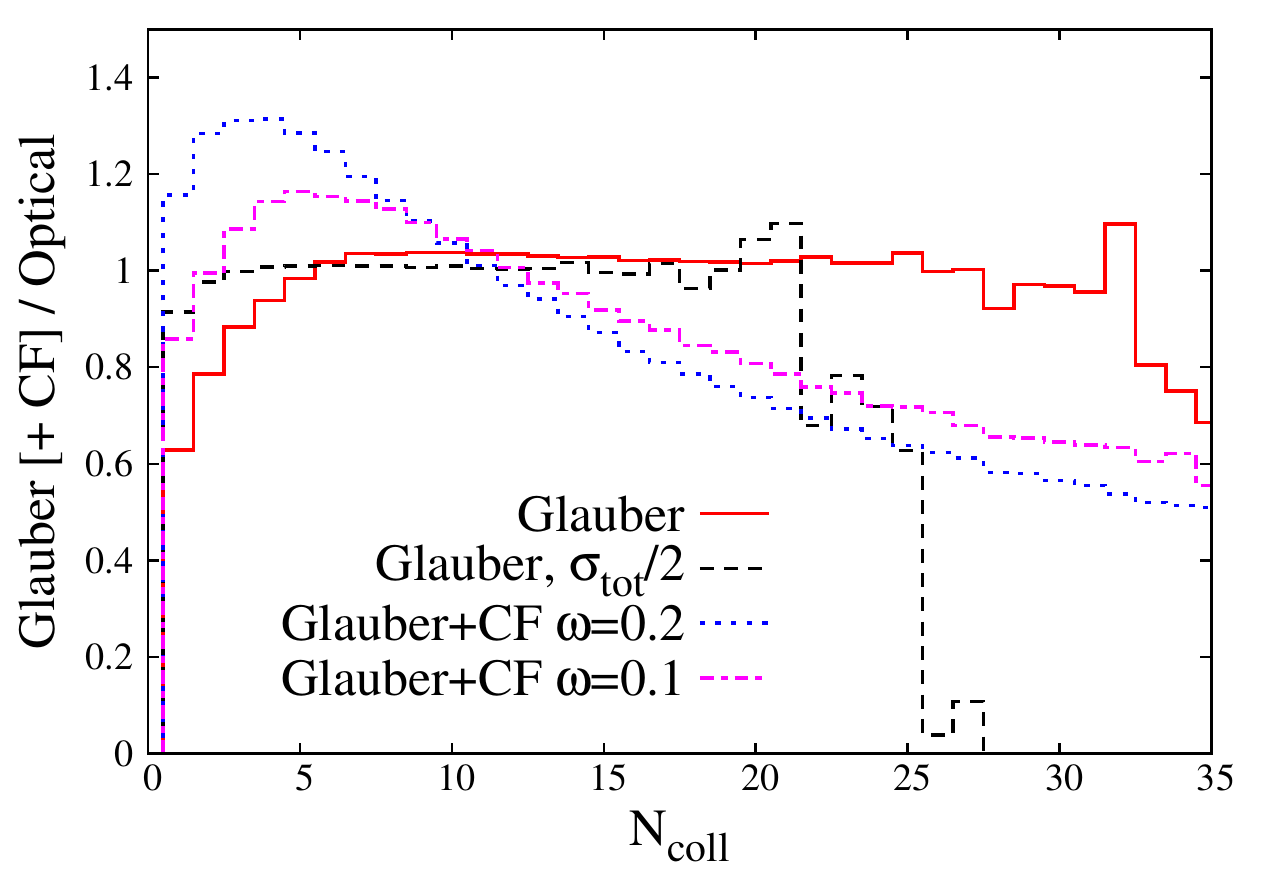}}
  \caption{Ratio $R_{HT}$ (Eq. (\ref{rnm})) of the rates of hard collisions in the Glauber 
    and the color fluctuation models  to that in  the optical model as a function of $N=N_{coll}$. }
  \label{rht}
\end{figure}

First, we consider the case of average  $x_p$ for which there is no significant correlation between the value of $\sigma$ for configuration and the parton distribution in the configuration. The case of $x_p$ for which such correlations maybe present in considered in the next section.
The results of our calculations are presented in Fig.~\ref{rht} for $\omega_{\sigma}=0$ (Glauber model) 
and for the CF model with $\omega_{\sigma}=0.1$ (our base model) and $\omega_{\sigma}=0.2$. Here we consider 
One can see that in the case of $\omega_{\sigma}=0$,  main deviations occur for small $N_{coll}$ and the effect 
decreases with a decrease of $\sigma_{tot}$. It appears that the main reason for this deviation is 
that the transverse gluon distribution in the nucleus is narrower than the soft interaction profile 
function  reflecting larger impact parameters in minimal bias $NN$ collisions than  those in hard $NN$ 
collisions \cite{Frankfurt:2003td, Frankfurt:2010ea}. As a result, at large impact parameters (small $N_{coll}$) the 
probability of hard collisions decreases as compared to the naive expectations. 
With a decrease of $\sigma_{tot}(pp)$  and, hence, the $b$-range of 
$NN$ interaction, the  deviation of $R_{HT}$ from unity is reduced. 

This effect was  first reported in \cite{Jia:2009mq} for $AA$ collisions at RHIC and the LHC and for d-Au collisions at 
RHIC energies using the  parameters of \cite{Frankfurt:2003td} for the impact parameter dependence  of hard 
collisions and a  simplified  model  for the  impact parameter dependence of $NN$ inelastic interactions.

Color fluctuations complicate the pattern 
of $N_{coll}$-dependence shown in Fig.~\ref{rht}
due to an additional effect of the broader distribution in $b$ of the collisions 
with small $\sigma$ (see Fig.~1 in \cite{Alvioli:2013vk}), which enhances the probability of collisions with small 
$N_{coll}$ for small impact parameters, where the parton transverse density is higher. 
At very large $N_{coll}$, yet another new effect takes place, namely, 
fluctuations with large $\sigma$ generate more collisions at large impact parameters, 
where the interaction is typically soft and does not lead to hard collisions. As a result, 
$R_{HT}$ becomes smaller than unity, while in the model without fluctuations, $R_{HT}$ 
stays very close to unity up to very large $N_{coll}$. 
We checked that results of our calculations are not sensitive to the presence of 
nucleon correlations in nuclei. 

As a result, the CF approach predicts a higher rate of events with a hard trigger starting at somewhat 
larger $N_{coll}$ than in minimum bias events (cf. Figs.~\ref{fig1} and \ref{rht1}).
Hence our analysis demonstrates that color fluctuations lead to the following two effects for 
large $N_{coll}$ for the bulk of hard observables: 
(i) the larger probability of collisions with $N_{coll} \ge 12$ and (ii) the reduced probability of hard 
subprocesses for the same $N_{coll}$ range. Further modeling is necessary to determine 
the optimal strategy to see these effects in the bulk data sample. 
Using the information on $x_p$ of the parton in the proton undergoing the
hard interaction may be an easier way forward.

\section{How to observe the effects of flickering in $pA$ collisions}
\label{sec:observ}

In this section  we  propose  strategies for using processes involving both soft and hard interactions 
to obtain the definitive evidence for the  presence of the flickering phenomenon.   The idea is to  investigate
 the correlation between the light-cone fraction 
 $x_p$ of the parton in the  proton involved in the hard collision 
and the  overall interaction strength of the configuration  containing this parton. 
The challenge for all such studies is that selection of certain classes of events 
(using a particular trigger) a priori  post-selects different configurations in both 
colliding systems and these two effects have to be disentangled. 

A natural question to ask is whether the parton distributions in configurations interacting 
with the strength smaller/larger than the average one are  different and whether  there 
exists a correlation between the presence of a parton with given $x$ (and virtuality) and 
the interaction strength of this configuration. Naively one should expect presence of such 
correlations at least for large $x$. Indeed, if we consider configurations with large $x$, e.g., 
$x >0.5$, one may expect that for such configurations the number of constituents should be 
smaller than on average (fewer $q\bar q$ pairs, etc.) as the consequence of the depletion of 
the phase volume for additional partons and selection of configuration with a minimal number 
of partons in the initial state before QCD evolution. Also, selection of $x$ much larger than 
the average one should select larger than average longitudinal and transverse  momenta in the  nucleon 
rest frame, leading to a smaller than average size, see, e.g., \cite{Coleman-Smith:2013rla,Frankfurt:1985cv}.
The shrinking may differ for large-$x$ $u$ and $d$ quarks since the $d/u$ ratio strongly depends on $x$
for $x\ge 0.4$, see \cite{Diehl:2013xca}.

Let us consider $pA$ collisions with a hard trigger which selects a parton with particular $x$ 
in the proton projectile. As in the inclusive case, we use the distribution over the number of wounded 
nucleons as in Eq.~(\ref{eq3}) with the substitution $P(\sigma)\to P(\sigma, x)$.  The distribution 
$P(\sigma, x)$ takes into account  the probability for a configuration with given $x$ to interact with the 
cross section $\sigma$. Due to the QCD evolution, $P(\sigma, x)$ also depends on the resolution scale 
(see Sect.~\ref{evolution}).  Let us 
suppose that one can roughly measure the effective number of interacting 
nucleons within the  nucleus, $N_{coll}$, based, e.g., on the energy release at the 
rapidities sufficiently far away from the central region 
(this is the strategy adopted by ATLAS~\cite{researchnote} and  CMS~\cite{CMSdijets}). 
  
We demonstrated in the previous section that deviations of $R_{HT}$ from unity are modest 
for fluctuations with 
$\sigma \le \sigma_{tot}/2$. Neglecting deviations of $R_{HT}$ from unity and nuclear modifications 
of PDFs (which is a small effect on the scale of the effects we consider here and which will 
be addressed later), we can use Eq.~(\ref{eq3}) to find the relation between  
$\left<\sigma(x) \right>$ and experimental  observables:
\beq 
    {\left< \sigma^2(x)\right> \over \sigma(x)} = { (\left<N_{coll} \right> -1 )  
      {A^2\over A-1} \over \int d^2b T^2(b)} \,.
    \label{maver}
\eeq
Similarly, we can use Eq.~(\ref{eq3}) to determine higher order moments of  $\sigma(x)$. 
For example, using Eq.~(\ref{eq3}) we find:
\beq 
    {\left< \sigma^3(x)\right> \over \sigma(x)} =  \left< (N_{coll}-2)(N_{coll}-1)\right> 
    {A^3  \over (A-1)(A-2) \int d^2b T^3(b)} \,.
    \label{maver2}
\end{equation}
Hence by combining  Eqs.~(\ref{maver}) and (\ref{maver2}) one can obtain information about  
the width  of the distribution over  $\sigma(x)$. 

A more accurate  calculation requires taking into account deviations from the $R_{HT}=1$ 
approximation used above which may be significant for large $N_{coll}$ (Sect.~\ref{hardsect}). 
Such an analysis would require much more elaborate modeling of  $pA$ collisions.

Another strategy is possible which allows one to amplify the effect of flickering. 
We can consider the distribution over $N_{coll}$ for $N_{coll}$ much larger 
than $\left<N_{coll}\right> $ for events with a hard trigger. 
In this case, scattering off small impact parameters dominates and fluctuations of $\sigma$ are enhanced relative 
to the fluctuations of the impact parameter
\footnote{In contrast, in $pp$ collisions, fluctuations of the impact parameter dominate in a wide kinematic range.  
For example, the strong positive correlation between the hadron multiplicity, 
$N_h$, and the rate of production of $J/\psi$, $D$, and $B$ mesons 
observed by ALICE \cite{Abelev:2012rz,Bala:2013yea} 
appears to be dominated by selection of different $b$ up to 
$N_h/\left<N_h\right > \sim 3$ \cite{Strikman:2011ar}.
The same pattern in the CMS data was recently demonstrated for high $p_t$ jet 
production~\cite{Azarkin:2014cja}.}.
  
For the reasons described above, we expect the strongest modification of the distribution 
over the number of collisions for large enough  $x_p $ 
(this automatically requires  large $p_t > $100 GeV/c for jets  for the 
current acceptance of the LHC detectors, which allows one to safely neglect leading twist nuclear shadowing
effects even if $x_A$ is small).  

\begin{figure}[!htp]
  \vskip -0.0cm
  \centerline{\hspace{0cm}
    \includegraphics[width=8.5cm]{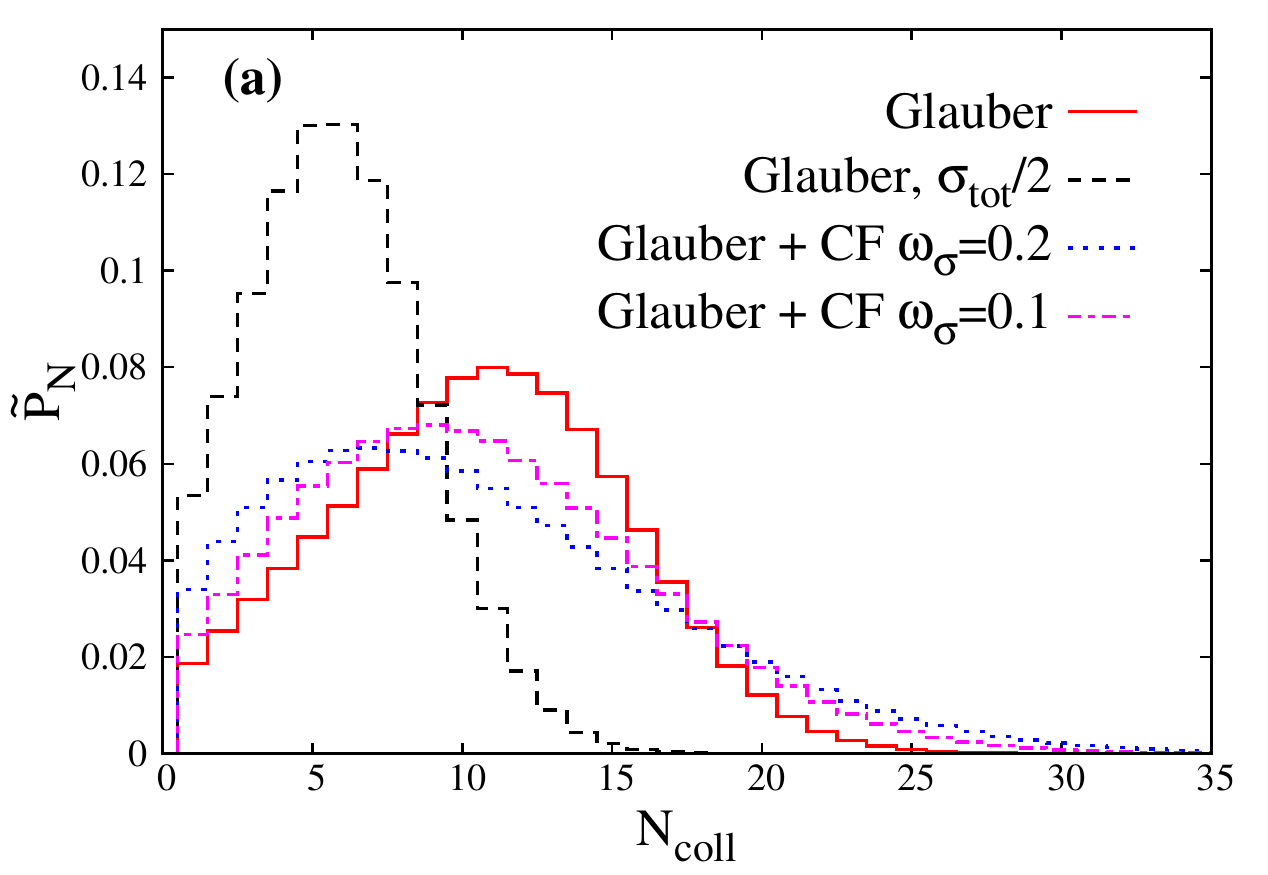}
    \hspace{0cm}\includegraphics[width=8.5cm]{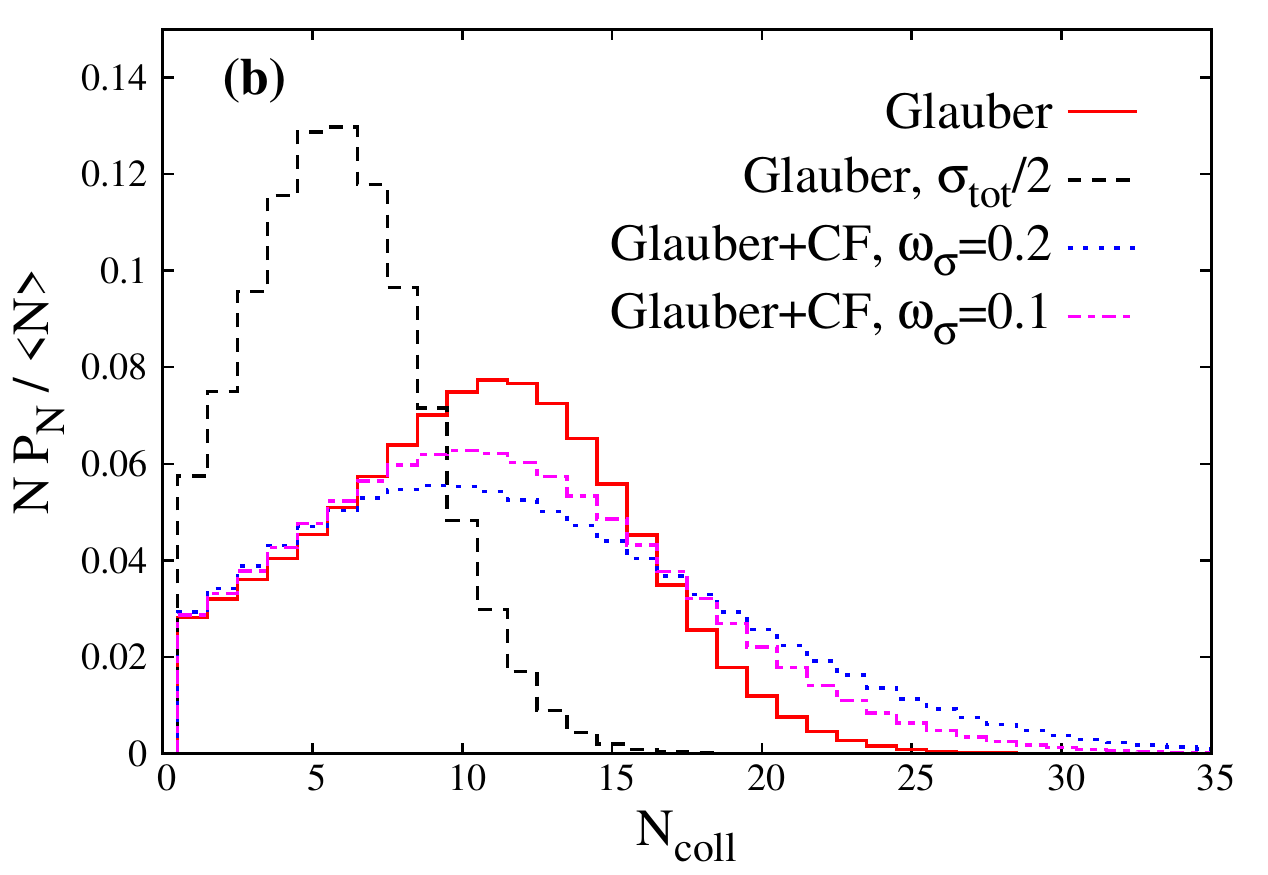}}
  \caption{The distribution over the number of collisions for a hard trigger 
    using (a) the full calculation and (b) the approximation $R_{HT}=1$.}
  \label{rht1}
\end{figure}

To study the sensitivity of the number of wounded nucleons to the average $\sigma(x)$ for configurations 
selected by the trigger, we performed calculations with $\left<\sigma\right>_x= \sigma_{tot}$, 
$\sigma_{tot}/2$ and $\sigma_{tot}/4$.  Within the CF picture, the following two effects compete 
in generating large $N_{coll}$ events: (i) selection of fluctuations in the nucleus wave function in which more 
nucleons happen to be at the impact parameter of the incoming proton (which, for  large $N_{coll}$ 
events, is anyway small $b < 3$ fm), and (ii) selection of fluctuations with $\sigma > \sigma(x)$. Our 
numerical studies show that there is large sensitivity to the mean value of $\sigma(x)$, 
even when  we allow for significant fluctuations of $\sigma(x)$. 

The results of these calculations are presented by the dashed curve in Fig.~\ref{rht1}.  One can see from the 
plot that for $N_{coll}$ larger than the average number of collisions $\left<N_{coll}\right> \approx 7$,  in the 
minimal bias events, one can easily observe the reduction of $\left<\sigma\right>_x$ by a factor of two.
To see whether flickering of the nucleon in the triggered  configuration can mimic the change of 
$\left<\sigma\right>_x$, we also considered the distributions for $\omega_{\sigma}=0.1$ and 0.2, see the dotted 
and dot-dashed curves in the figure.  One can see from the figure that this effect is not large enough to prevent 
the observation of reduction of $\left<\sigma\right>_x$.  The opposite limit is that of small enough $x_p$. In this 
case one would trigger on configurations with $\left<\sigma\right>$  larger than the average one 
leading to broadening of 
the distribution over $N_{coll}$. 

\begin{figure}[!htp]
  \vskip -0.0cm
  \centerline{\hspace{0cm}
    \includegraphics[width=8.5cm]{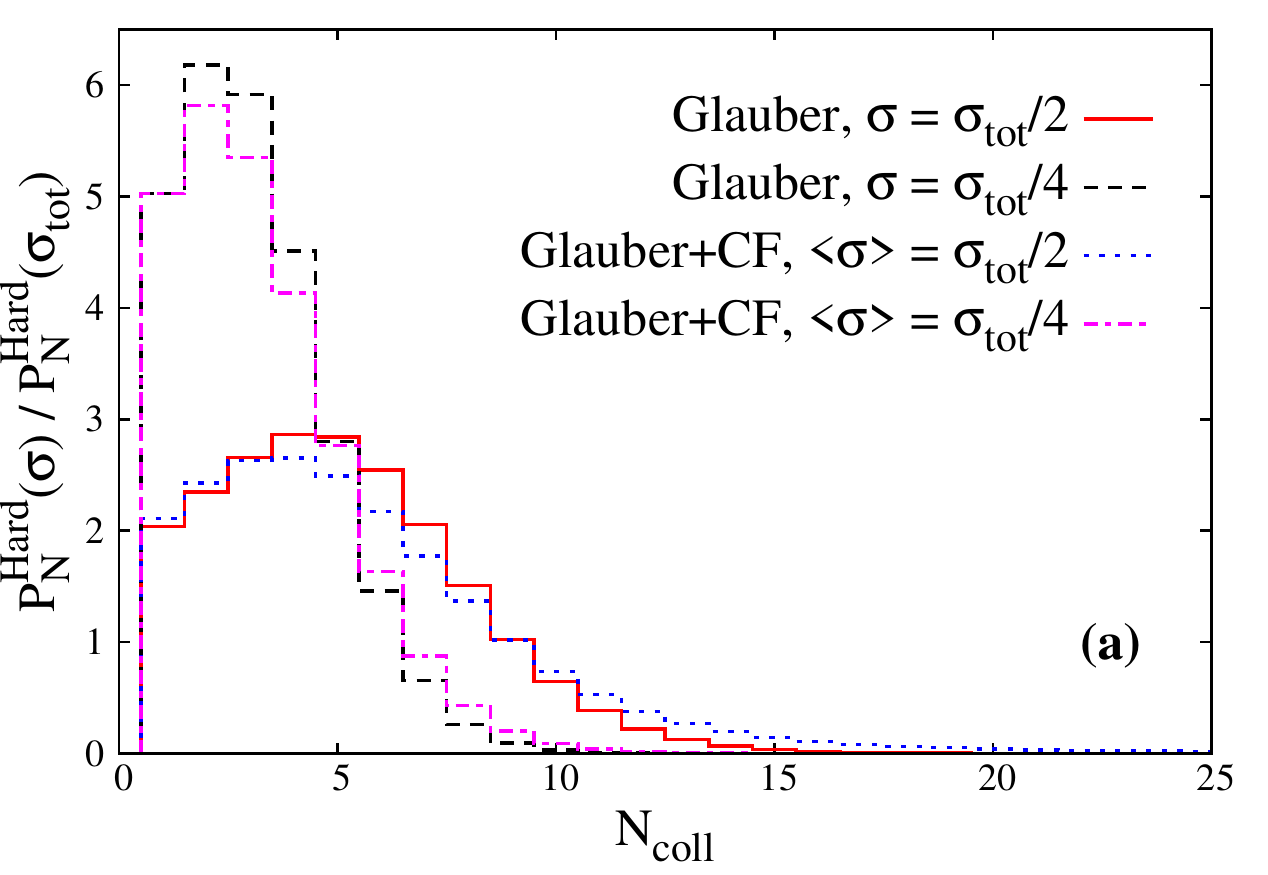}
    \hspace{0cm}\includegraphics[width=8.5cm]{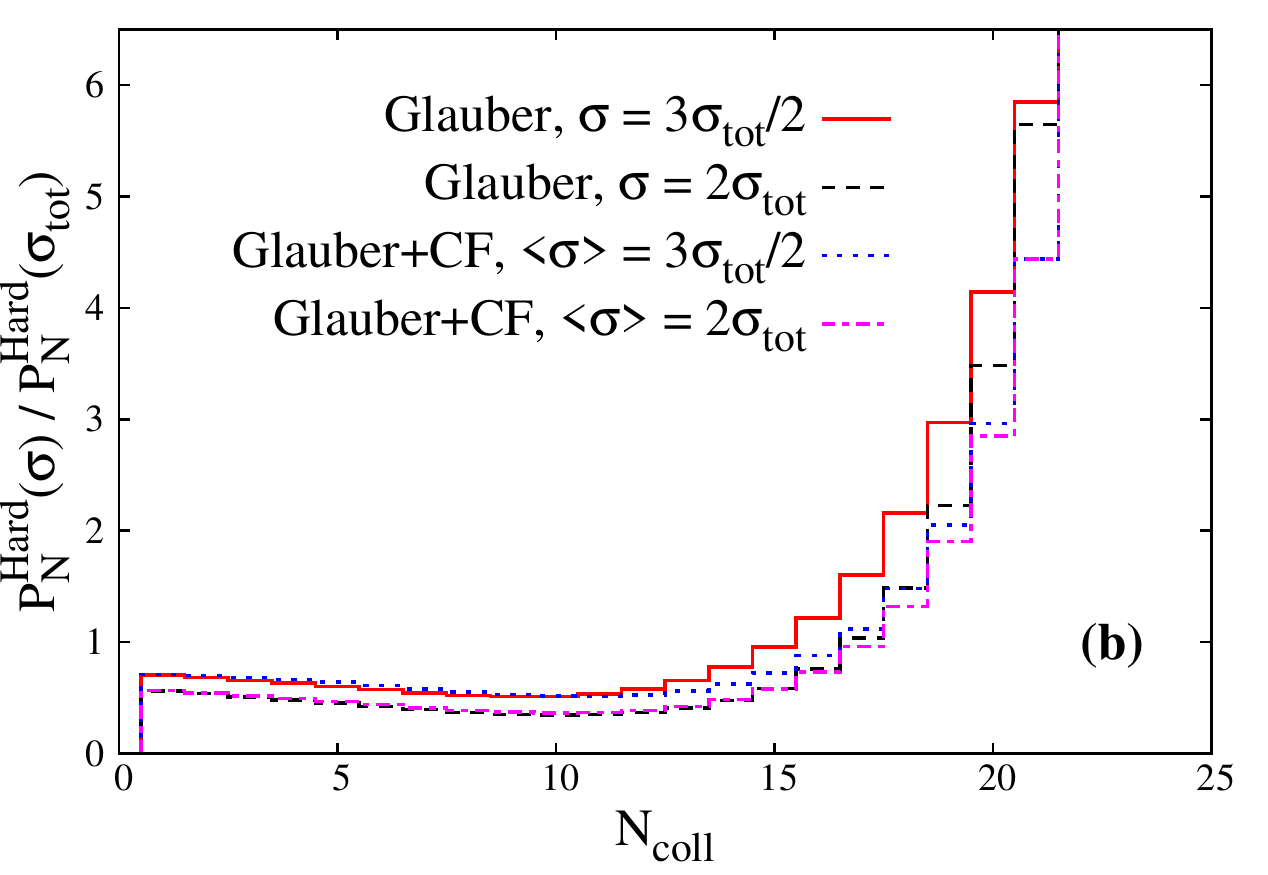}}
  \centerline{\hspace{0cm}
    \includegraphics[width=8.5cm]{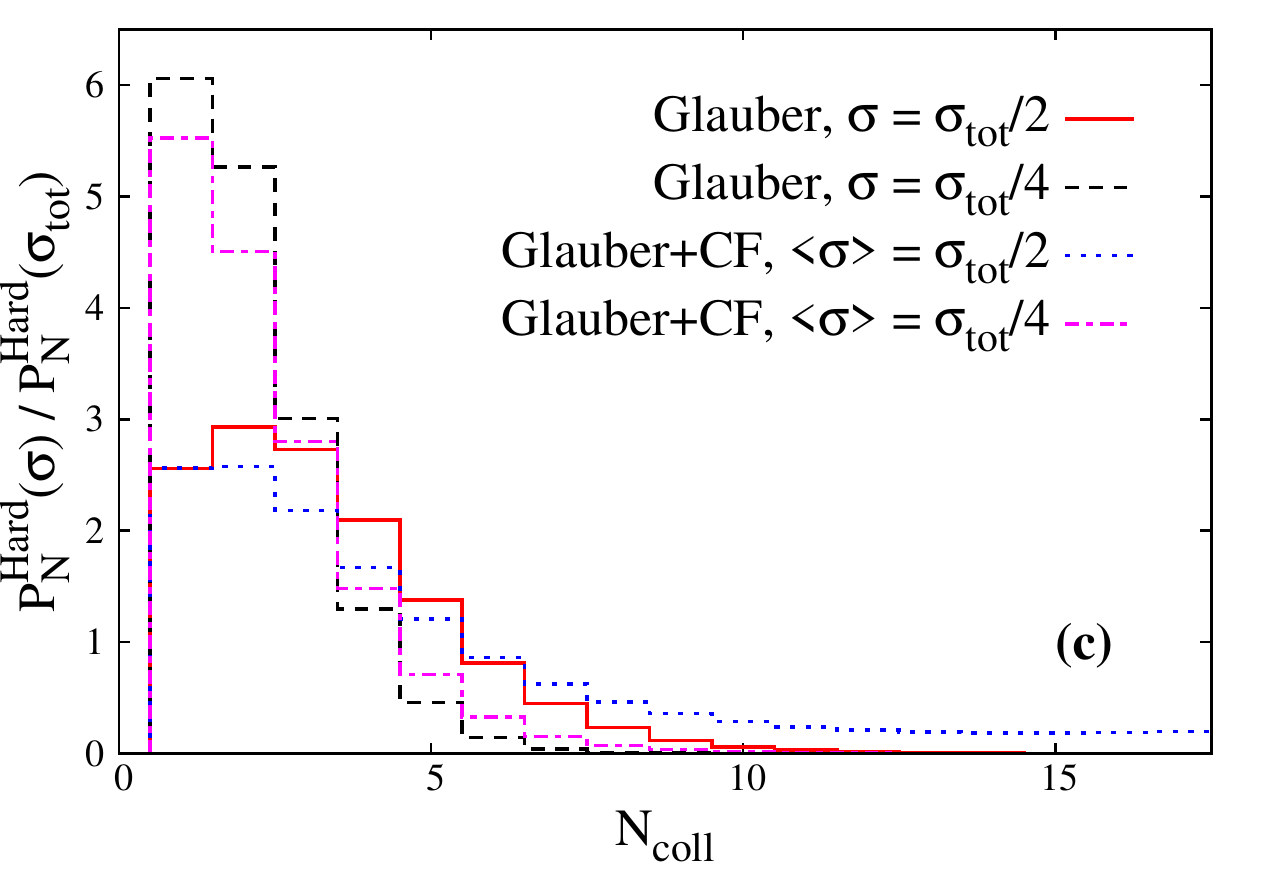}
    \hspace{0cm}\includegraphics[width=8.5cm]{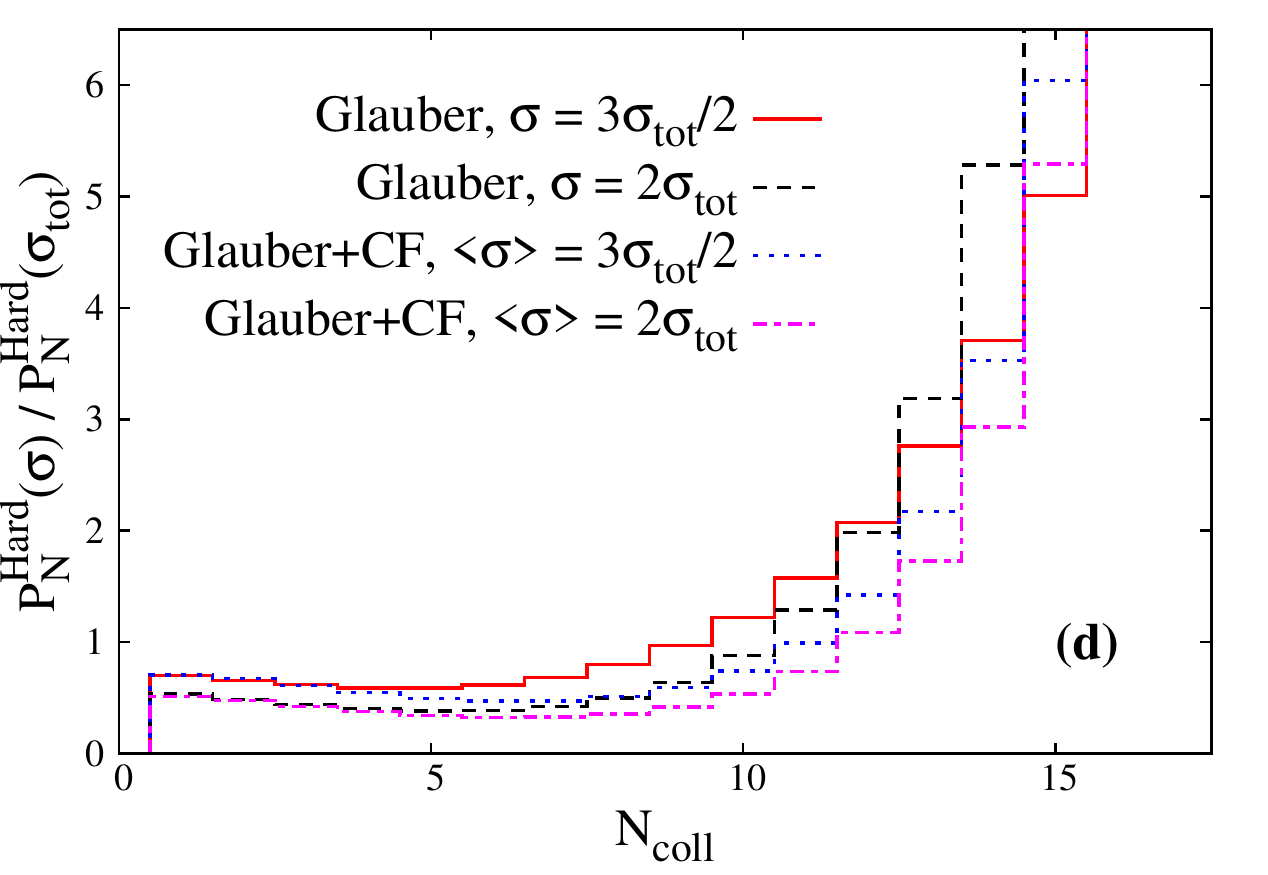}}
  \caption{Ratio of the probabilities $P_N$ of having $N=N_{coll}$ wounded nucleons for 
    configurations with different $\left<\sigma(x)\right>$ and $P_N$ for $\sigma=\sigma_{tot}$
    at LHC (panels (a) and (b)) and RHIC (panels (c) and (d)) energies.
    The ratio is averaged over the global impact parameter $b$ and plotted as a function of 
    $N=N_{coll}$. The solid and dashed curves neglect the dispersion 
    of $\sigma$, while the dotted and dot-dashed curves show the results obtained with a Gaussian 
    distribution around $\left<\sigma(x)\right>$ with the variance equal to 0.1. Panels (a) and (c)
    show results for $NN$ interaction cross sections smaller than average, while panels (b) and (d)
    show results for $NN$ interaction cross sections larger than average.
}
  \label{figSIXbis}
\end{figure}

To illustrate the possible magnitude of the change in the $x_A$ distribution as a function of 
$N_{coll}$, we present in Fig.~\ref{figSIXbis} the ratios of $P_N(\sigma(x))/P_N(\sigma=  \sigma_{in})$ 
for $\sigma(x) /\sigma_{in}=2$, 1.5, 0.5, and 0.25 and 
$\omega_{\sigma}=0$ and $\omega_{\sigma}=0.1$ (for LHC energies) and $\omega_{\sigma}=0.25$ (for RHIC energies) calculated using 
the procedure of Sect.~\ref{hardsect}.

\begin{figure}[!htp]
  \vskip -0.0cm
  \centerline{\includegraphics[width=8.5cm]{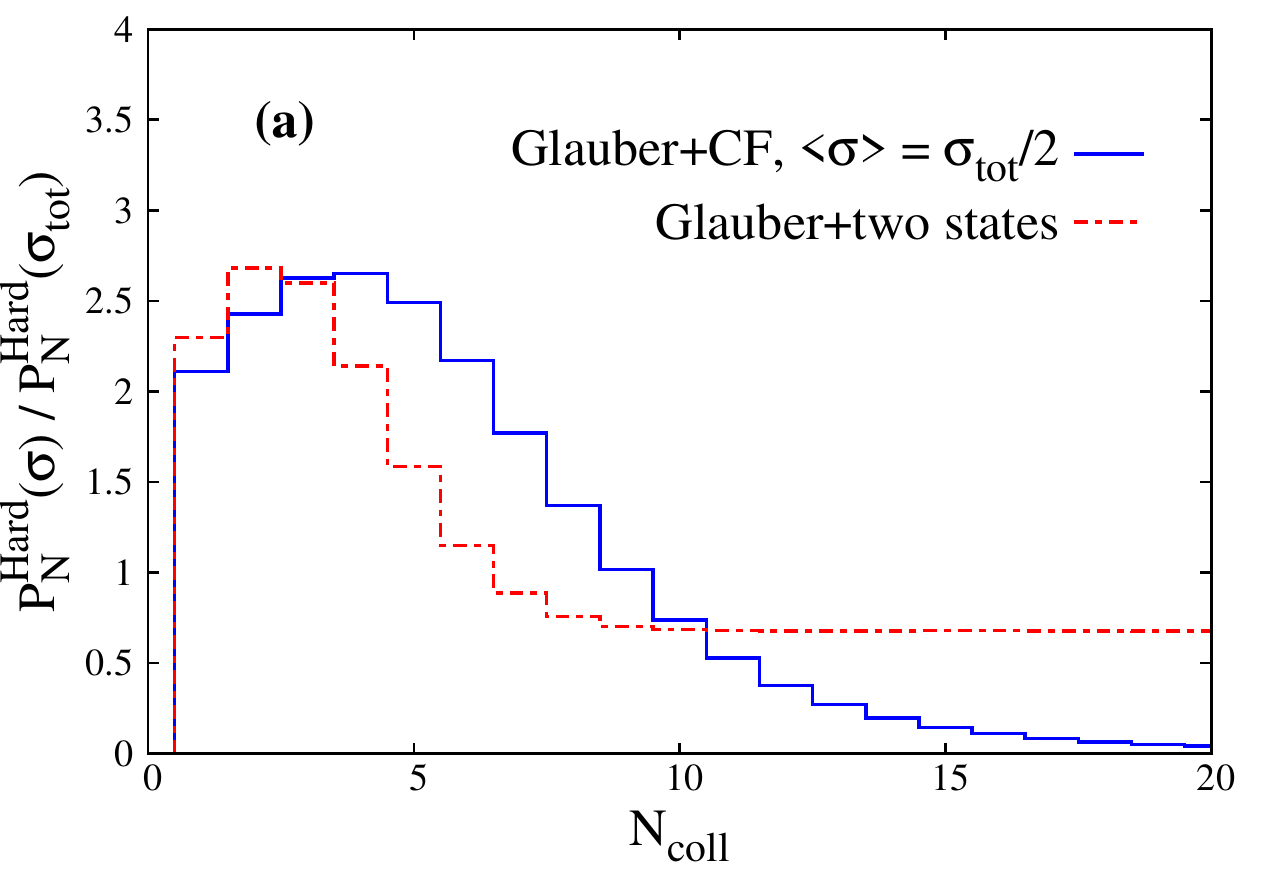}
    \includegraphics[width=8.5cm]{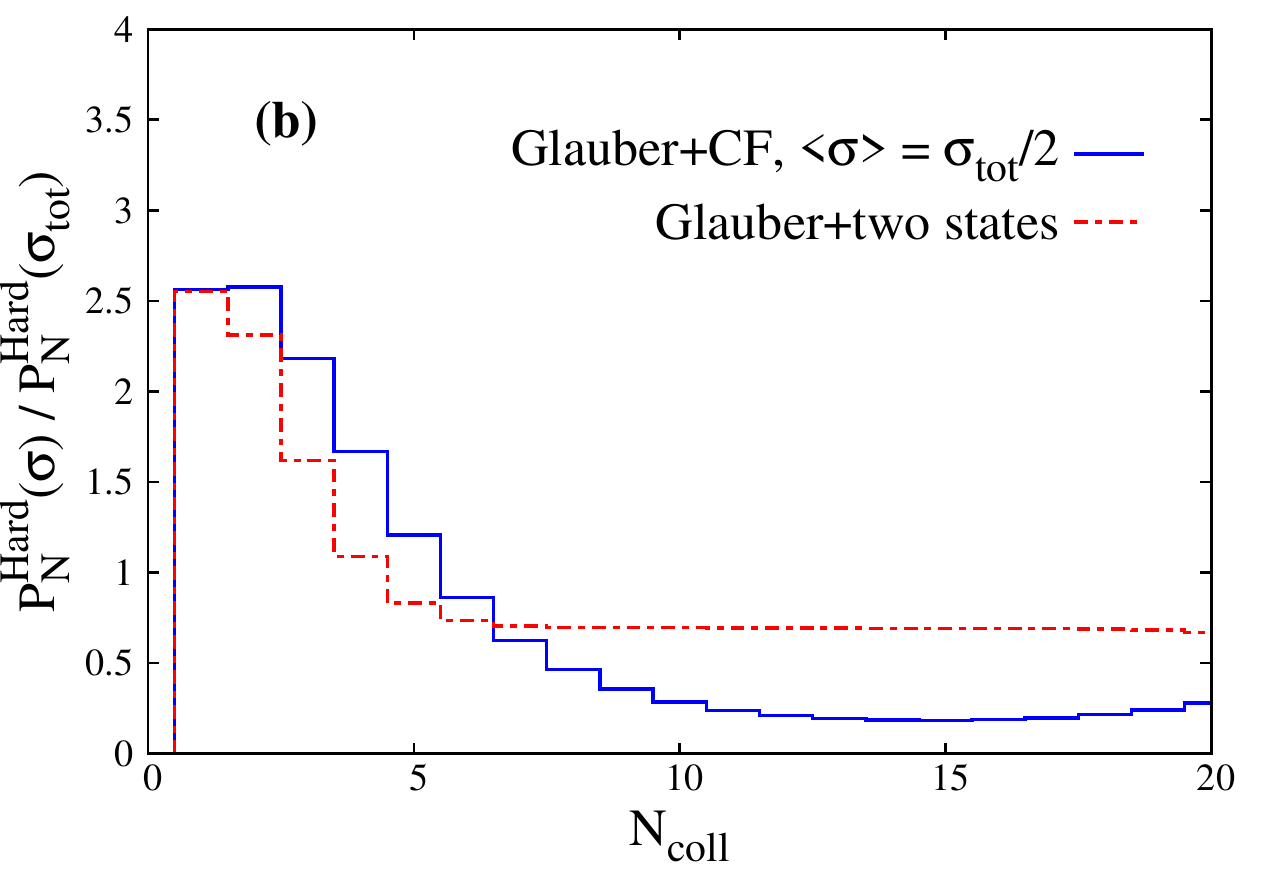}}
  \caption{Ratio of the probabilities $P_N$ of having $N=N_{coll}$ wounded nucleons for 
    configurations with different $\left<\sigma(x)\right>$ and $P_N$ for $\sigma=\sigma_{tot}$
    for LHC (a) and RHIC (b) energies. The ratio is averaged over the global impact parameter 
    $b$ and plotted as a function of $N=N_{coll}$. The dashed curve is also shown in Fig. 
    \ref{figSIXbis}; see text for the definition of the two-state model shown by the dotted 
    curve.
  }
  \label{figSIX}
\end{figure}

To illustrate the sensitivity to the pattern of flickering for fixed $x$, 
we use the scenario  where  $\left<\sigma(x)\right>=\sigma_{tot}/2$ and 
proton fluctuations consist of 
two states with probabilities 2/3 and 1/3 with the respective cross sections $\sigma_{tot}/4$ and $\sigma_{tot}$.
We compare the results of this model and the Gaussian-like model with 
the same variance equal 1/2   in Fig. \ref{figSIX}. One can see that 
deviations from the results of the calculation with $\sigma=\sigma_{tot}$ are large in both cases .
There is also significant difference in the high-$N_{coll}$ tail.

Note in passing that the best way to check the difference between the transverse sizes of 
configurations with leading $u$- and $d$-quarks would be to measure leading $W^+$ and $W^-$ 
production (one additional advantage is that in this case energy conservation effects would be 
the same for the two channels). Similarly, one can look for the difference in the accompanying  
multiplicity for forward $W^{\pm}$ production in $pp$ scattering \cite{Frankfurt:2004kn}.

Overall an inspection of the numerical results presented in Figs.~\ref{rht1}, \ref{figSIXbis}, and  \ref{figSIX}
indicates that 
the selection of events with the highest nuclear activity---for example, the top 1\%---greatly 
amplifies effects of flickering. 
Namely, the relative contribution of events with small $\sigma$ is suppressed much stronger 
than in the events with smaller nuclear activity, leading to a strong distortion 
of the dijet distribution over $x_p$. 
Large-$x_p$ rates (which are dominated by scattering off valence quarks of the proton) 
should be suppressed, while small-$x$ rates, which are dominated by scattering off gluons,
should be enhanced. A complementary way 
to study this effect is to consider the distribution over the energy deposited in the 
calorimeter as a function of $x_p$. 
We expect the monotonous  shrinkage of the distribution over the number of collisions with 
increasing $x$, with the 
strongest effect for the highest number of collisions. 
Note in passing that such a study  allows one to test the conjecture 
that large-$x$ triggers select significantly smaller than the average-size configurations 
in the nucleon. Hence, such a study 
would allow one to rule out/confirm the explanation of the EMC effect as being 
due to the suppression of small-size configurations 
in bound nucleons \cite{Frankfurt:1985cv}.

The discussed patterns do not depend on details of the relation between $N_{coll}$ and 
the signal in the calorimeter at negative rapidities (in the direction of the nucleus fragmentation).   
Qualitatively the discussed pattern is consistent with the pattern reported by ATLAS \cite{researchnote} and 
CMS \cite{CMSdijets}. Indeed,  ATLAS observes the suppression of production of leading jets which they find to 
be predominantly a function of  $x_p$, while the CMS   analysis presents the correlation of the  
calorimeter activity with a  different  quantity $(\eta_{Jet_1} + \eta_{Jet_2})/2$ 
which still reflects the value of $x_p$~
\footnote{Importance  of small-size configurations 
at large $x_p$ could also be studied in hard diffraction at the 
LHC by studying hard diffraction at fixed $x_{\Pomeron}$ and fixed $\beta$ 
(the fraction of the energy carried by the parton 
belonging to the diffracting proton) as a function of $x_p$. The gap survival probability 
should increase when $x_p \ge 0.5$.}.

In order to perform a detailed  comparison of the CF model with the LHC data one needs data in bins of $x_p$. A preliminary version of such data was   presented so far by ATLAS only. Also one needs  
 a realistic model/models for the distribution 
over $E_T$ for events with given $N_{coll}$. Such an analysis is underway and will be presented elsewhere. 
At the same time, we can obtain an estimate of the magnitude of the necessary change of average $\sigma(x\sim 0.5) $ using the data for most peripheral collisions ( 90\%--60\% centrality) where the expected enhancement is a rather weak function of $N_{coll}$. The data indicate an enhancement of the jet rate by a factor of about two. This corresponds to $\sigma(x\sim 0.5) \sim \sigma_{tot}/2 $. It is worth emphasizing here that presence of an enhancement would be difficult to understand based on the logic of energy losses.

\section{Perturbative QCD evolution of  $P(\sigma,x)$}
\label{evolution}

The distribution $P(\sigma,x)$ characterizes the distribution of strength of soft interactions of the configuration containing a parton carrying the 
light-cone fraction $x$  at a sufficiently small resolution scale. 
A change of the scale---e.g., a change of $p_T$ of the jets---does not change the strength of the soft interaction
but   reduces $x$ of the parton.  Hence,  one  can deduce 
an evolution equation for $P_i(\sigma,x)$  expressing $P_i(\sigma,x)$ at the large scale $Q^2$ through 
$P_i(\sigma,x)$ at the input $Q_0^2$ scale ($i=q,\bar q,g$). For $x \ge 0.2 $, where we expect  a significant 
dependence of  $P_{q(g)}(\sigma, x, Q_0^2)$ on $x$,  perturbative QCD (pQCD) evolution leads to 
a decrease of     $\sigma(x, Q^2)$  with an increase of $Q^2$.  
 This is because the account of the QCD radiation---$Q^2$ evolution---shows that partons with given $x$ and 
 large 
 $Q^2$ originate from 
 larger
 $x$  at  the nonperturbative scale $Q_0^2$.  As we argued above, for large $x$, the size of configuration 
 is likely to decrease with an increase of $x$. Hence, the increase of $p_t$ of the trigger for fixed $x$ 
 should lead to a gradual decrease of the average $\sigma$ for the dominant configurations. 
  In addition, in the gluon channel,   one also expects a significant mixing between the contributions of 
 (anti) quarks and gluons at $Q_0^2$.
 
\begin{figure}[htp]
  \vskip -0.0cm
  \centerline{\includegraphics[width=16.0cm]{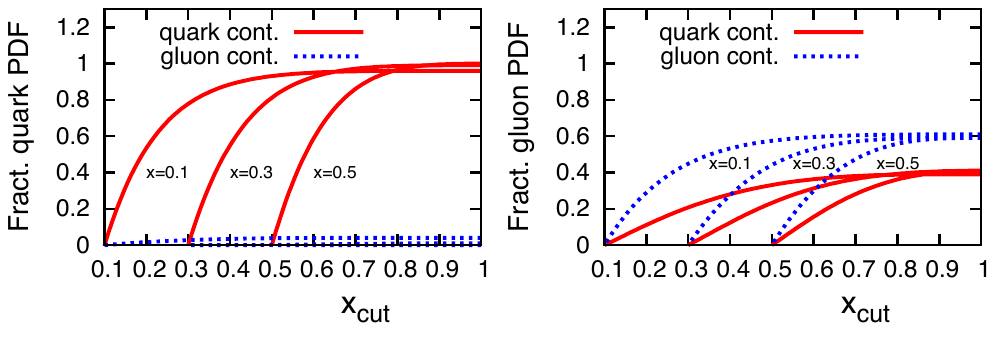}}
  \caption{Fractional contributions to the quark (left panel) and gluon (right panel) PDFs at $x=0.1$, 0.3, and 0.5 and 
  $Q^2= 10^4$ GeV$^2$ originating from the interval $[x,x_{\rm cut}]$ at the input scale $Q_0^2= 4$ GeV$^2$.
  The solid curves correspond to the quark contribution and the dotted curves are for the gluon contribution. }
  \label{curve}
\end{figure}

 To illustrate these effects, we used  the Dokshitzer--Gribov--Lipatov--Altarelli--Parisi
 (DGLAP) evolution equations to evaluate the contributions of quark and gluon PDFs at $Q_0^2= 4$ GeV$^2$
to the quark and gluon PDFs at $Q^2= 10^4$ GeV$^2$. 
The results of this analysis are presented in Fig.~\ref{curve} as fractions of the parton distribution 
(the left panel is for the $u$-quark PDF and the right panel is for the gluon PDF)  at 
given $x$ ($x=0.1$, 0.3, and 0.5) and $Q^2= 10^4$ GeV$^2$, which originate from the quark (solid curves) and 
gluon (dotted curves) PDFs at the input scale 
$Q_0^2= 4$ GeV$^2$, which have the support on the $[x,x_{\rm cut}]$ interval.
The plotted fractions are shown as functions of the cut-off parameter $x_{\rm cut}$,  $x_{\rm cut}> x$.
Thus, by construction, the shown fractions vanish in the $x_{\rm cut} \to x$ limit and rapidly tend to unity
in the $x_{\rm cut} \to 1$ limit. 
Varying the parameter $x_{\rm cut}$ we examine the weight of different intervals of the light-cone variable 
$x^{\prime}$ at the input scale of the DGLAP evolution in the resulting quark and gluon PDFs at the higher scale $Q^2$.
Such an analysis allows one to quantitatively study the effective trajectory of QCD evolution. (For an analysis of 
QCD evolution trajectories at small $x$, see~\cite{Frankfurt:2011cs}).

 One can see from the figure that
(i)  $x_{\rm cut}$ is noticeably larger than $x$ which means that the PDFs at high $Q^2$ originate from the broad 
$[x,x_{\rm cut}]$ interval at the input scale, and
(ii) the gluon PDF receives a significant though not dominant contribution also from quarks at
the initial scale. This effect is somewhat smaller for lower $Q^2$. 
In summary, Fig.~\ref{curve} illustrates that
perturbative QCD evolution induces fluctuations  in $\sigma$ even if there is no dispersion at the initial scale.

\section{Fluctuations and conditional parton distributions}
\label{7}

In the previous sections we made a simplifying approximation that nuclear PDFs are the sums of nucleon PDFs. 
 Deviations from this approximation are observed at $x\ge 0.4$ (the EMC effect) and small $x$. 
In the long run it would be possible to use the discussed processes to also study novel aspects of the nucleus partonic  
structure since they select nuclear configurations 
where many more nucleons are located in the cylinder around the transverse 
position of the hard interaction than on  average. 
These high density nuclear configurations should have the different parton 
structure for at least for two reasons:  
(i) the leading twist nuclear shadowing 
should increase
with a decrease of $x$  due to an increase of the nucleon density 
in the  cylinder of target nucleons interacting with the projectile 
at a fixed impact parameter  as progressively more nucleons screen each other;  (ii) the decrease of average internucleon distances 
within the cylinder should increase the modification of large-$x$ parton  distributions, i.e., 
the EMC effect, which is roughly proportional to the probability of the short 
range correlations in nuclei~\cite{Frankfurt:1985cv,Hen:2013oha}. 

\subsection{Leading twist nuclear shadowing effect}

In Sect.~\ref{hardsect} we calculated the dependence of the nuclear gluon density 
(treated as a sum of the nucleon gluon densities)   encountered by the 
projectile parton as a function of $N_{coll}$. We have demonstrated that the pattern strongly depends on the 
strength of fluctuations: if the fluctuations are neglected, 
the density is to a very good approximation given by $N_{coll}\,  g_N(x, Q^2)$. At the same 
time, fluctuations slow down the increase of the gluon distribution
by the factor of $R_{HT}$ presented in Fig.~\ref{rht}.

Qualitatively we expect that with an increase of $N_{coll}$,   nuclear 
shadowing  for small $x_A < 0.01$ and antishadowing for  $x_A\sim 0.1$ will increase. 
In the following,   we will use the 
theory of leading twist nuclear shadowing, see the review in \cite{Frankfurt:2011cs}, to calculate the shadowing and 
compensating antishadowing effects. We restrict ourself to the limit when $x_p$ of the parton of the proton is small enough ($\le 0.2$) so that we can use $P_N(\sigma)$.

As a reference point, we consider the ratio of nuclear PDFs at the zero impact parameter 
$g_A(x,Q^2, b=0) $ and the properly normalized nucleon gluon density:
\beq
\frac{g_A(x,Q^2, b=0)}{T_A(b=0)g_N(x,Q^2)} \,,
\eeq
which was calculated in Section 5.5 of \cite{Frankfurt:2011cs}.

The effective transverse gluon density probed by the projectile is: 
\begin{equation}
g_{A}(x, Q^2, N_{coll})
=  {N_{coll}R_{HT}(N_{coll}) \over N_{coll}(b=0) R_{HT}(N_{coll}(b=0)) }g_A(x,Q^2, b=0)\,.
\end{equation}
Defining now the ratio of the effective gluon densities for $N_{coll}$ as 
\begin{equation}
  k=  {N_{coll}R_{HT}(N_{coll}) \over N_{coll}(b=0) R_{HT}(N_{coll}(b=0))} \,, 
\end{equation}
we can calculate the shadowing and antishadowing effects---to a good approximation---by rescaling  the nuclear density 
in the equations determining  the shadowing effect by the factor of $k$. 
Using the results of Sec.~\ref{sec:fluct}, we find 
$\left <N_{coll}\right> \approx 14.5$ and from the inspection of Fig.~\ref{rht} one can 
see that $k$ can reach for large $N_{coll} $ the values of up to $k=2$.

\begin{figure}[htp]
  \centerline{\includegraphics[width=8.5cm]{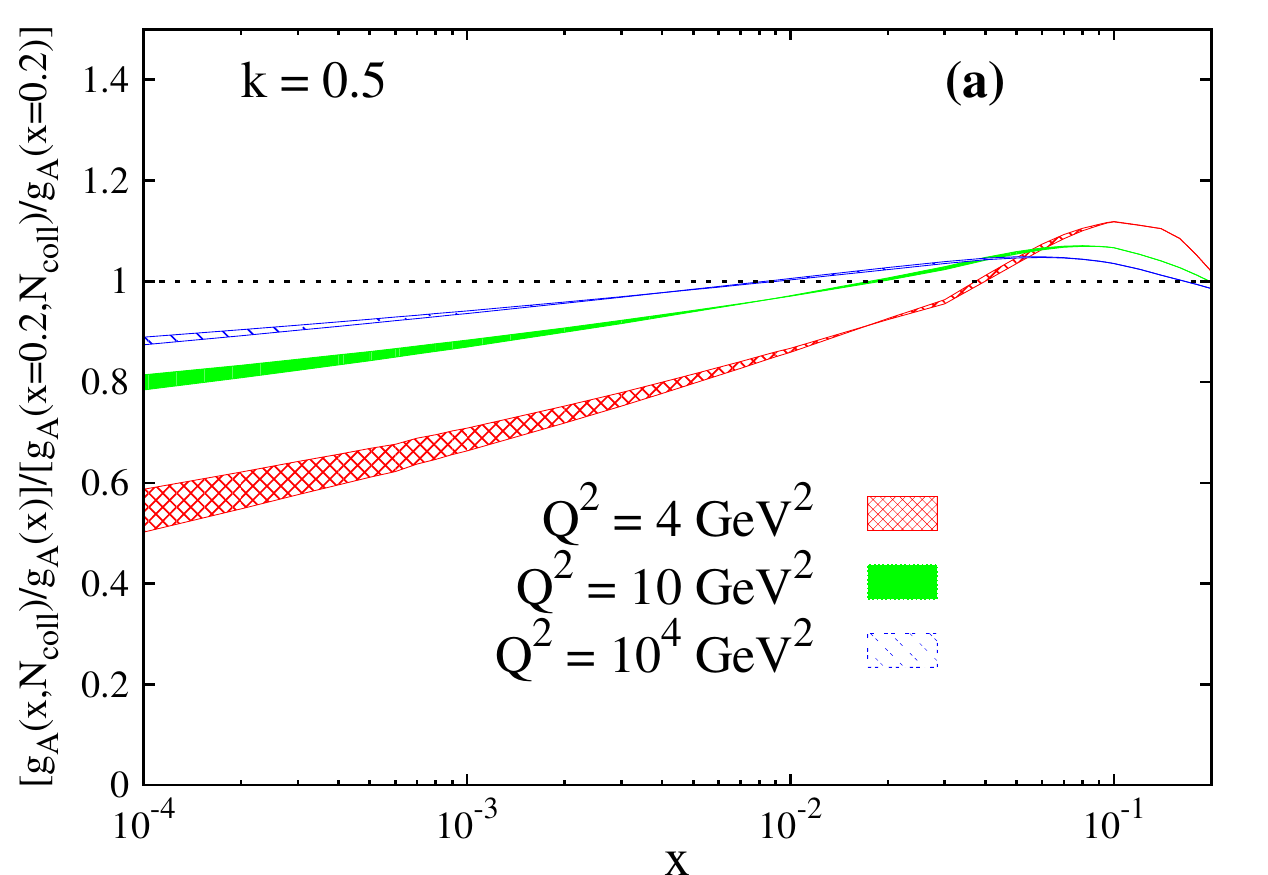}
    \hspace{0cm}\includegraphics[width=8.5cm]{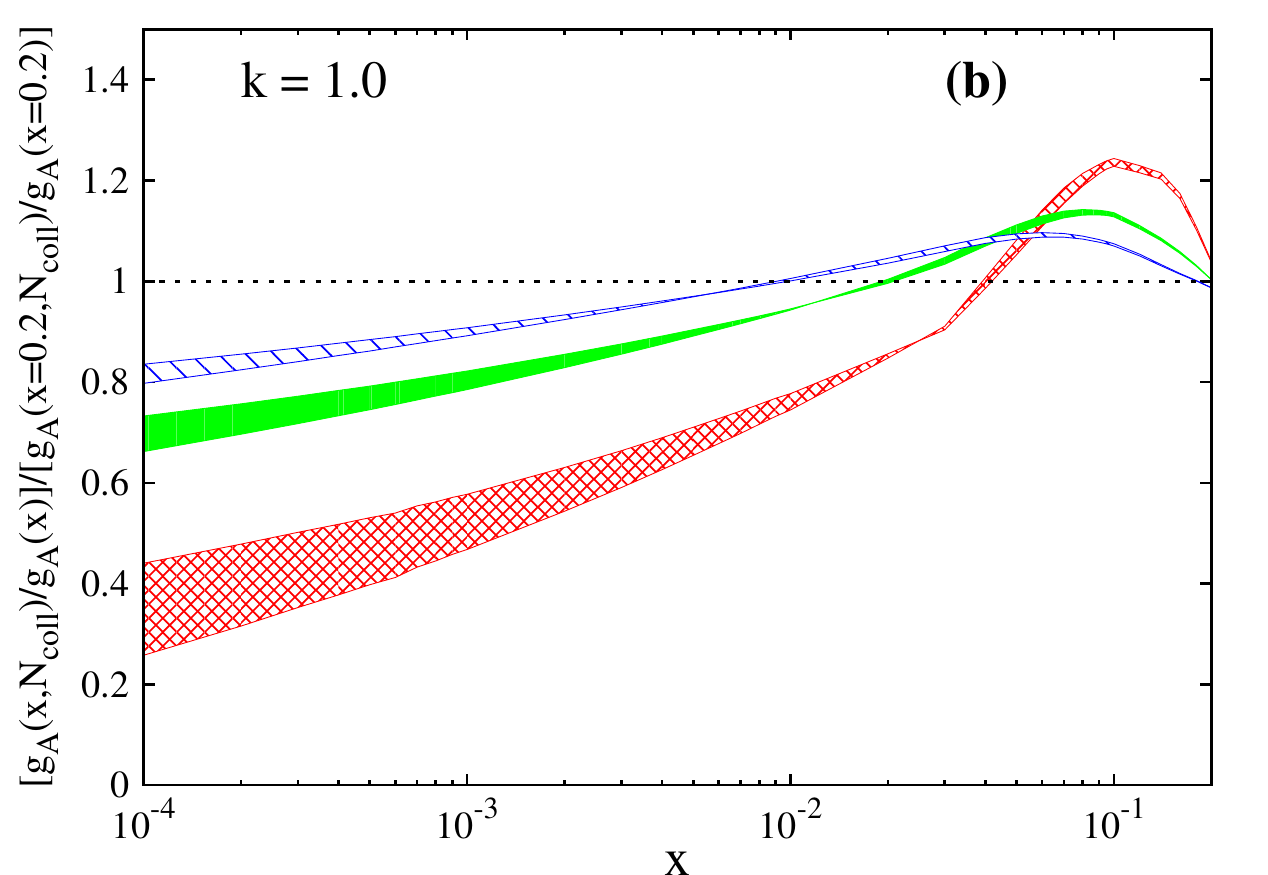}}
  \centerline{\includegraphics[width=8.5cm]{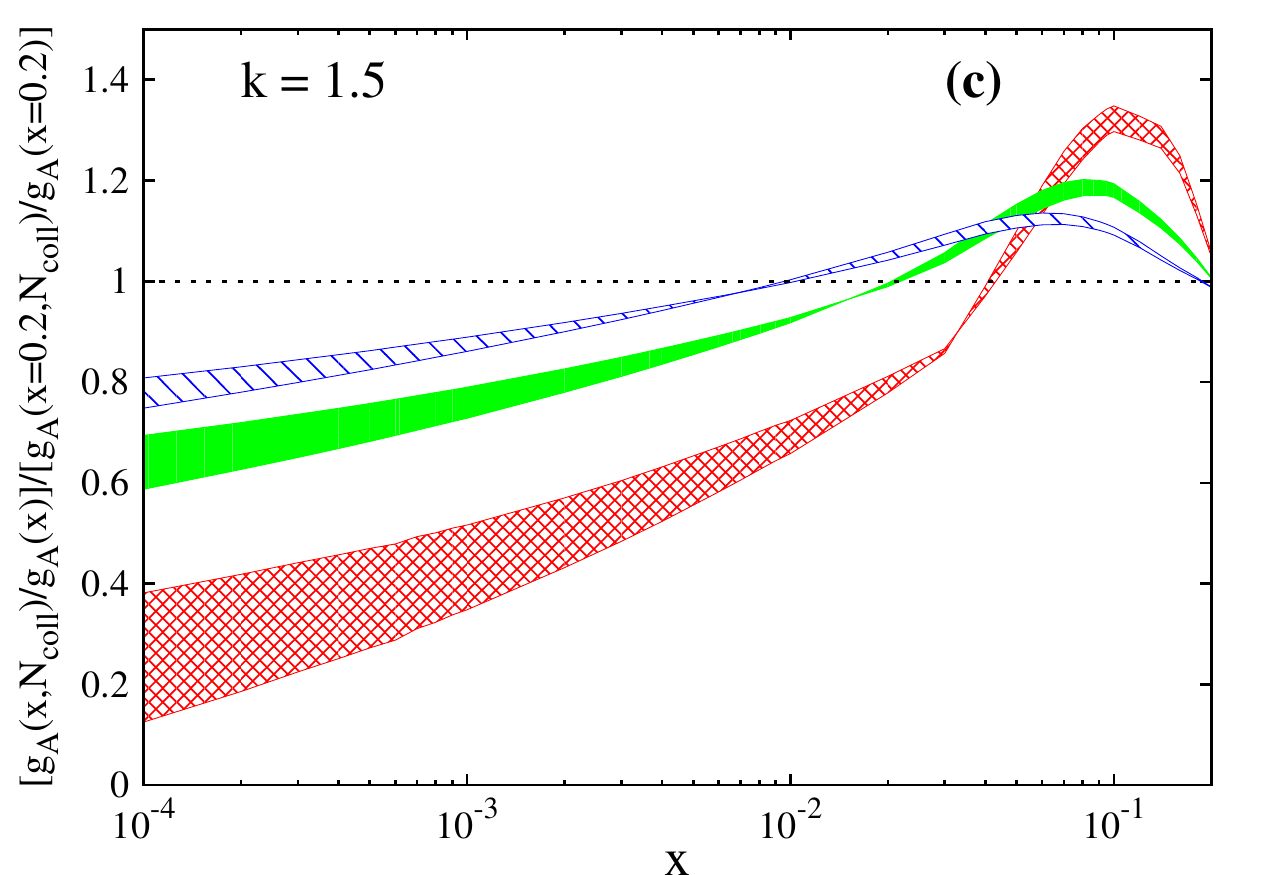}
    \hspace{0cm}\includegraphics[width=8.5cm]{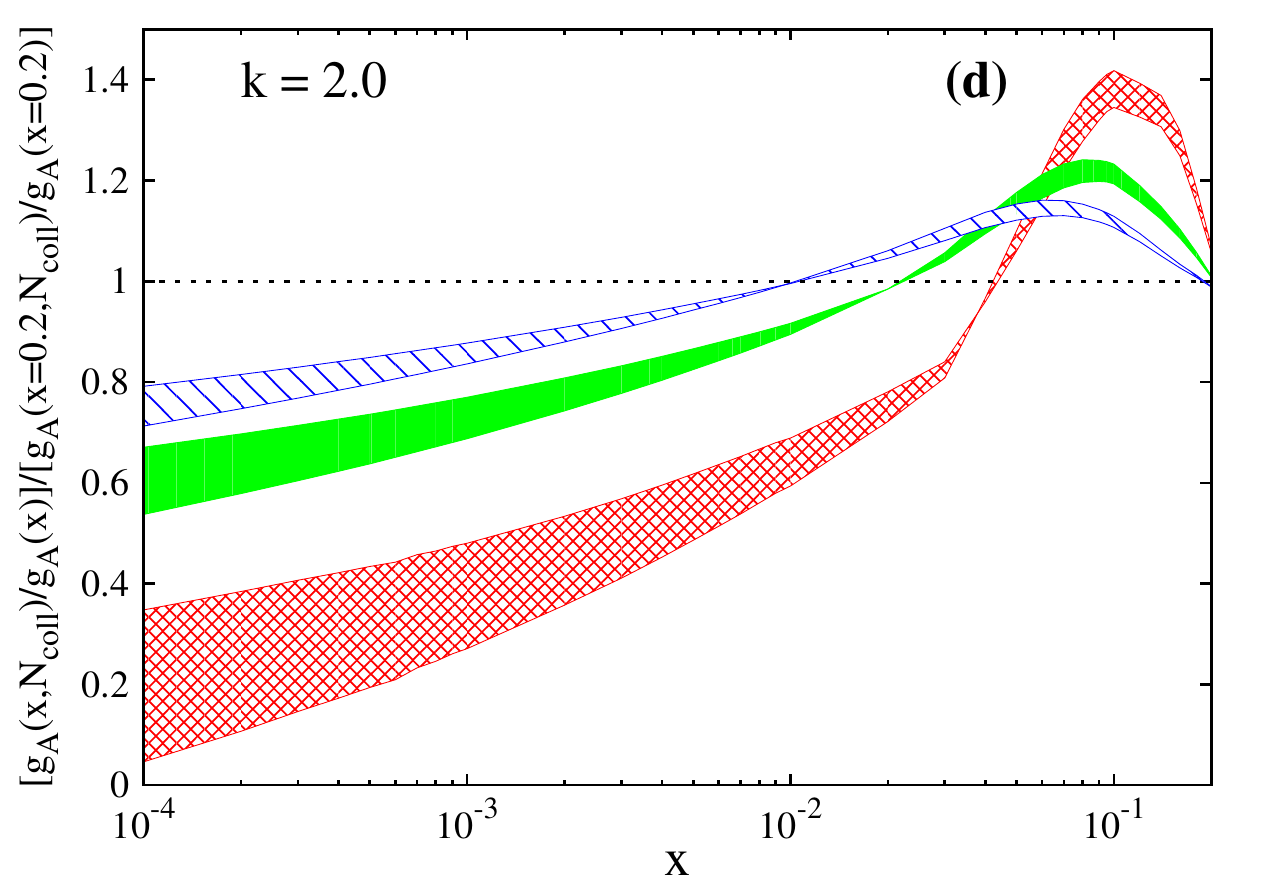}}
  \caption{Ratio of the  nuclear 
  gluon conditional distribution
  for given  $N_{coll} $ and the inclusive gluon 
    density normalized to their values at $x= 0.2$
    as a function of $x$ for different values of $Q^2$ and $k$.
    See text for details.}
   \label{figX}
\end{figure}

Figure~\ref{figX} presents our predictions for the super ratio of $(g_{A}(x, Q^2,N_{coll})/g_A(x,Q^2))/g_{A}(x=0.2, Q^2,N_{coll})/g_A(x=0.2,Q^2))$ as a function of $x$ for three values of $Q^2=4$, 10, and 10$^4$ GeV$^2$ and four
values of $k=0.5$, 1., 1.5, and 2.
The shaded bands represent the theoretical uncertainty of the leading twist theory of nuclear shadowing associated 
with modeling of multiple (three and more) interactions of a hard probe with a nucleus~\cite{Frankfurt:2011cs}.
One can see from the figure that the expected modifications of nuclear conditional PDFs 
should be rather large, 
if one could use a hard probe with a moderate virtuality of, e.g., 100 GeV$^2$. 
For the case of dijets with $p_t\ge 100$ GeV/c, the effect is rather
small for a wide range of $x$ and represents a small correction for the  
 studies of the effects of selection of large $x_p$ in the currently studied processes 
with a dijet trigger.

Note here that due to uncertainties in the procedure for determination of $N_{coll}$, the
optimal procedure would be to consider the ratios of cross sections for small $x_A$ and 
$x_A\sim 0.2$, where nuclear effects are  negligibly small, for the 
same $N_{coll}$ and to preferably use the same range of $x_p$.
 
The average $N_{coll}$ for the top 1\% of collisions can be estimated  using the results 
presented in Fig.~1. We find for 
these collisions that  $\left<N_{coll}\right> \sim 20  (25) $ for $\omega_{\sigma}=0 (0.1)$ and, 
hence, $k\sim$ 1.5 (1.25), which corresponds  
to quite a significant deviation from the $x$ dependence of inclusive nuclear PDFs.

\begin{figure}[htp]
  \vskip -0.0cm
  \centerline{\includegraphics[width=15.0cm]{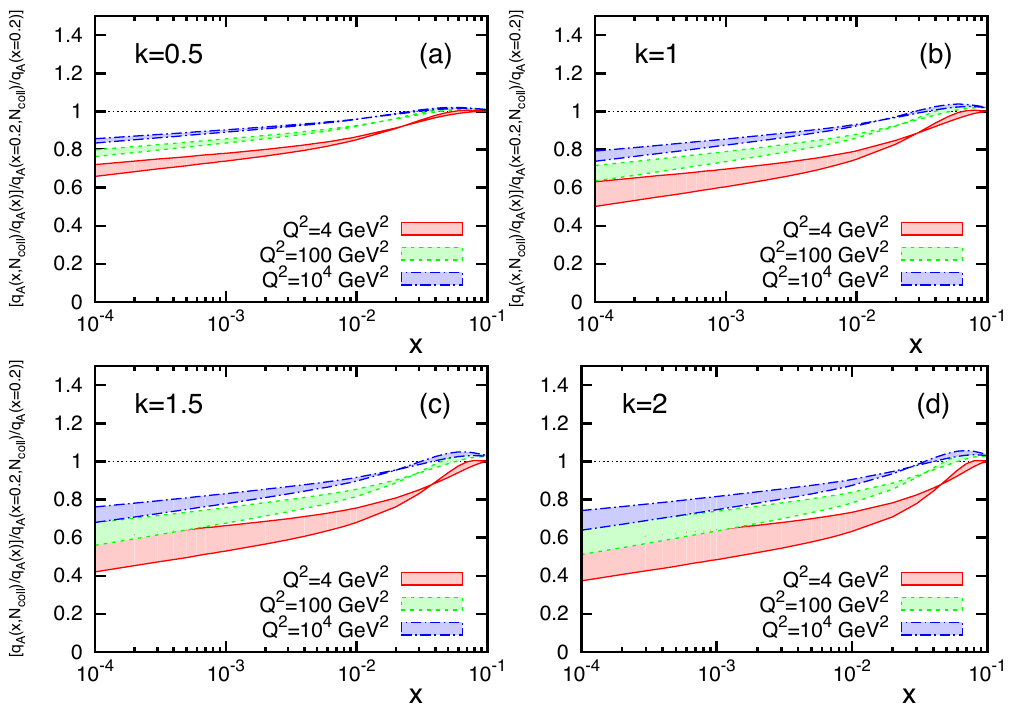}}
  \caption{The ${\bar u}_{A}$ quark superratio $({\bar u}_{A}(x, Q^2,N_{coll})/{\bar u}_A(x,Q^2))/{\bar u}_{A}(x=0.2, Q^2,N_{coll})/{\bar u}_A(x=0.2,Q^2))$ as a function of $x$ for different values of $Q^2$ and $k$.
  See Fig.~\ref{figX} for comparison and text for details.}
  
  \label{figX1}
\end{figure}

The quark channel analogue of Fig.~\ref{figX},
Fig.~\ref{figX1} shows our predictions for the superratio $({\bar u}_{A}(x, Q^2,N_{coll})/{\bar u}_A(x,Q^2))/{\bar u}_{A}(x=0.2, Q^2,N_{coll})/{\bar u}_A(x=0.2,Q^2))$ for the ${\bar u}_{A}$ quark.

\subsection{The $x_A\sim 0.5 $ region}

The above calculation demonstrates that the distances between nucleons in nuclei
are reduced for large-$N_{coll}$ triggers. 
This should have implications for the large-$x_A$ conditional PDFs of the nucleus. 
Indeed it is known that nuclear PDFs at large $x_A$ are significantly suppressed 
as compared to the free nucleon ones for $x$ between 0.5 and 0.7 and 
large $Q^2$. The scale of the suppression for heavy nuclei and large $Q^2$ is on the scale 
of 20\% as measured at CERN 
in the kinematics where the leading twist definitely dominates, see the review~\cite{Arneodo:1992wf}.

It is natural to expect that the EMC effect originates due to pairs of nucleons 
coming close together and deforming  
each other wave functions. 
Higher the nucleon momentum, further it is off-mass-shell, and, hence, larger the effect is.  
Hence one can expect that 
the EMC effect
 is mostly due to the presence of 
short-range correlations  \cite{Frankfurt:1985cv}. 
The recent analyses of the data are consistent with this expectation, see the 
review and references in \cite{Hen:2013oha}.

For heavy nuclei, the probability of short-range correlations (SRCs) is approximately 
proportional to the local nuclear density.  Hence, one can estimate 
the magnitude of the modification of nuclear PDFs due to selection of the large-$N_{coll}$ events as 
\beq 
    {(1- f_{A}/f_{N})_{N_{coll}} \over 1- f_{A}/f_{N}} \sim k \,,
\eeq
where $f_A$ and $f_N$ denote the quark nucleus and nucleon densities, respectively.
Since $k \sim 1.3 - 1.5 $ for the 1\% of events with the highest $N_{coll}$, 
the expected change of the EMC effect is rather modest. 
Still this selection appears to provide a unique  opportunity to probe nuclear matter at 
the density significantly 
higher than the average one.  
 
A more accurate analysis  should take into account the 
dominance of $pn$ correlations, see review 
in \cite{Frankfurt:2008zv,Alvioli:2013qyz}, 
the interplay between attraction and repulsion in SRCs, etc. 
Such an analysis will be presented elsewhere.

\section{Suggestions for  future analyses}
\label{sec:future}

In the future analyses of the data it would be important to study jet production as a function of centrality  for bins of $x_p$ 
and $x_A$ to separate possible effects of the conditional nuclear PDFs and effects of color fluctuations.  Testing that 
different processes with the same $x_p$ show the same centrality pattern is critical.

 It would be also interesting to study the effect at fixed $x_p$ as a function of $p_t$ of the jet. Such a dependence arises 
 due to DGLAP evolution since $\sigma$ for a configuration  depends on the ''parent'' $x_p$ at the low $Q^2$ scale, 
 which is larger than $x_p$ for the jet  (Fig.~\ref{curve}). 
  
Studies of fluctuation effects for  $P_g(x,Q^2)$  in the gluon channel will be challenging as the  deviations from average  
due to squeezing are expected for  $x >  0.2 -  0.3$ at the input scale $Q_0^2$  corresponding to somewhat smaller $x$ 
for jets with $p_T\sim 100$ GeV (Fig.~\ref{curve}). 
Still the crossover point between the gluon and quark contributions for such $p_T$ is $x\sim 0.2$ so that in view of 
the significant quark contribution to $g_N(x,Q^2)$, the effect of the smaller average 
gluon  $\sigma_g(x,Q_0^2) $ would be rather small --- on the scale of 30\%.
  Hence one would need to use the processes where the gluon contribution is enhanced, for example,  production of 
  heavy quarks  at relatively modest $p_t$, which is obviously experimentally challenging.
Nevertheless it  would be highly desirable to study CF effects separately for quarks and gluons  since 
the squeezing is likely to be different and starts in the gluon case at smaller $x$.

 If one would observe a pattern similar to the one for generic jets,  it  would strongly suggest presence of the EMC effect for gluons due to suppression of weakly interacting contributions
  in bound nucleons \cite{Frankfurt:1985cv}.

One should also look for kinematics of small $x_p$ where the contribution of configurations with $\sigma$ larger than average should be enhanced.

Measurements using $W^{\pm}$ can be used to study the difference of the interaction strength of configurations with leading  
$u$- and $d$-quarks. An advantage of this process which maybe possible to study at RHIC in the forward kinematics is that any effects related to energy--momentum conservation  cancel out in the ratio of the cross sections at same $x$.

\section{Conclusions}
\label{sec:conclusions}
 
In conclusion, we have demonstrated that it is possible to use the LHC $pA$ data to 
understand the correlation between the parton 
distribution in the nucleon and its interaction strength and to explore fine details 
of the nuclear parton structure in the EMC effect and nuclear shadowing
regions.

The authors thank members of the ALICE, ATLAS and CMS collaborations
and especially B.~Cole, D.~C.~Gulhan, Y.-J.~Lee, and 
J.~Schukraft for useful discussions. M.~Strikman's research was supported  by the U.S. Department of Energy Office of Science, Office of Nuclear Physics under Award Number DE-FG02-93ER40771. L.~Frankfurt research was supported by BSF grant.

\appendix
\section{On the account of the momentum conservation in the color fluctuation approach}

Conservation of momentum implies that the proton momentum in proton--nucleus inelastic collisions is split 
among  several collisions. Hence, the energy released in one inelastic $pN$ collision is a decreasing function 
of the  number of collisions  and 
it is necessity to take this effect into account. The aim of this appendix is to  explain 
that   energy--momentum conservation is effectively  taken into account 
in  the color fluctuation (CF) formulae for the total cross sections,  the number of wounded nucleons, etc.   
In contrast, the celebrated  Abramovsky--Gribov--Kancheli (AGK) cancelation \cite{AGK}  
among shadowing contributions for the single inclusive spectrum including 
inelastic processes due to the cut of $N\ge 2$ ladders for central rapidities is violated and the 
resulting formulae 
contain the  additional factor $R_{N_{coll}}$ which cannot  be evaluated at present in a model-independent way,
see Eq.~(\ref{em}).  The explanation of above statements  involves several steps which  are outlined below.

In QCD,  longitudinal distances  comparable to the atomic scale dominate in $pA$ collisions
at the LHC (to simplify 
the  discussion, we work in the nucleus rest frame). Indeed it follows from the uncertainty principle that the lifetime of a  
fast proton with the momentum $P_N$ and the energy $E$   in the configuration $\left|n\right>$ is:
\begin{equation}
t_{coh} = {1\over (E_{n}-E)}  = {2P_N \over \sum_i {m_i^2 + p_{i\,t}^2\over x_i}-m_{p}^2} \,,
\label{coh}
\end{equation}
where $m_i$, $x_i$, and $p_{i\,t}$ are the masses of constituents, their light-cone fractions and transverse momenta, 
respectively.  Hence, during the passage through the nucleus and far behind it, the transverse positions  of the fast 
constituents of the projectile do not change.  These constituents interact with a target through ladders attached to these 
constituents. This interaction may destroy coherence of these constituents with spectator constituents leading to 
multi-hadron production. 

It follows from Eq.~(\ref{coh})   that the proton energy is divided among fast   partons long before the collision.
So  the  energy--momentum  conservation is explicitly satisfied for the interaction of partons with a target. On the contrary, 
in the Glauber picture, the projectile nucleon is destroyed in the first collision and combines  back into the nucleon during 
the time between  collisions with different nucleons of the nucleus. This is obviously impossible at high energies since  
such a transition takes too long a time   $\approx t_{coh}$.  Another problem is  that due to energy--momentum conservation, a 
significant part of the projectile energy is  already lost  in the first inelastic collision diminishing the 
phase volume for other 
inelastic collisions.  The Glauber model  derived within quantum mechanics ignores energy--momentum conservation  which 
is  controversial when $N\ge 2$ ladders are cut.  These puzzles  are naturally resolved in QCD  since the contribution of the 
planar Feynman diagrams relevant for the Glauber model disappears at high energies where processes with hadron 
production dominate. The complete cancellation of the planar diagrams has been demonstrated for 
high energy processes  by direct calculations of the relevant Feynman diagrams  in Refs.~\cite{Mandelstam,PomeronCalculus}
using analytic properties of amplitudes in the plane of masses of diffractively produced states. 

Gribov  suggested to decompose the contribution of non-planar diagrams over  the sum of the pole corresponding to the 
initial hadron and inelastic diffractive states. 
Exploring both representations---kind of duality between quark--gluon and hadron degrees of freedom---allows one
 to 
analyze implications of the energy--momentum conservation. In practice the derived formulae for nuclear shadowing 
differ from  the formulae of the Glauber approximation by    the  small inelastic 
shadowing correction~\cite{Gribov}.  This pre-QCD approach leads to the  following models : 
(i) the Gribov--Glauber model, which includes inelastic diffractive processes in the intermediate states, and 
(ii) the  color fluctuation approach~\cite{Heiselberg:1991is,Alvioli:2013vk}, which takes into account the 
fluctuations of the interaction strength in the form familiar from the properties of bound states in QCD. 

The color fluctuation approach~\cite{Heiselberg:1991is} is a generalization of  the pre-QCD assumption of 
Good and Walker \cite{Good:1960ba} that one can present the high energy hadron--nucleus interaction as 
a superposition of interactions of the initial hadron in the configurations of different strengths which do not 
change during the propagation through the nucleus.  The CF  approach includes low-mass fluctuations as well
as the fluctuations into large diffractive masses.    The natural pattern for the contribution of large diffractive masses   
is the triple Pomeron mechanism which takes into account that the intermediate masses  increase  with energy.  This 
mechanism allows for the splitting  of energy in the interaction with several nucleons to occur at rapidities rather far 
away from the nucleon's rapidity providing a mechanism for production of leading nucleons in the interactions of the 
proton with  several nucleons.  

In the case of the hadron interaction with two nucleons, the  shadowing correction to the total cross section is expressed 
through the cross section of diffraction (elastic plus  inelastic)~\cite{Gribov}.  This Gribov formula follows also from 
the Abramovsky--Gribov--Kancheli combinatorics for cross sections~\cite{AGK}.     It  follows also 
from the model~\cite{Kopeliovich:1978qz},  which includes
fluctuations of the interaction strength in the form of   the Miettinen--Pumplin relation [Eq.~(\ref{diffr1})].  
The Gribov formula for shadowing in  proton - deuteron scattering  includes the triple Pomeron contribution exactly 
and allows one to express the shadowing contribution to $\sigma_{tot}(pd)$ through the diffractive cut of the Feynman diagrams with exchange by two ladders. 
 So for the interaction with two nucleons, energy--momentum conservation is accurately taken into account. 
Higher moments are also expressed through experimental observable, see the determination of 
$\left<\sigma^3\right> $ in \cite{Heiselberg:1991is}. 

 Thus, all factors related to the increase of the cross section with energy  are included into $P(\sigma)$. 
 No additional factors in the CF formulae are required to describe also the number of wounded nucleons  since   
 it is  evaluated through the multiplicity of hadrons in the kinematics close to the nucleus fragmentation region~\cite{Atlas}.  
 In this kinematics hadron multiplicity is a slow   function of $s$  as the consequence of the Feynman scaling.

For hadron multiplicity in the center of rapidity and in the proton fragmentation region, the answer is  more complicated. 
Note here that  the hadron  inclusive cross section at central rapidities  in  $pp$ interaction grows with energy approximately 
as $(s/s_0)^\kappa$, where  $\kappa \sim 0.25$.  Thus,  the  hadron inclusive cross section for the $pN$ interaction contains  
the factor of $(x_{i}s/s_0)^\kappa$
instead of $(s/s_0)^\kappa$ within the Gribov--Glauber model and the CF approach,  where $x_i$ is the fraction of the projectile momentum carried by the interacting parton "i" 
or a group of partons. The factor $(x_i)^{\kappa}$ is not included  in $P(\sigma)$  since it is  additional to 
the CF series in terms of $\left<\sigma^n\right> $  defining $P(\sigma)$. Hence,   it follows from energy--momentum 
conservation   that  the hadron inclusive cross section due to the processes  where  $N>1$  ladders are cut is suppressed by 
the factor of $R_N$ as compared to the formulae of the Gribov--Glauber approximation and the CF approach combined with the 
AGK cutting rules: 
\beq
R_{N}= \frac{\sum_n \int d \tau_n \left|\psi_{n}(x_{1},...x_{N},...x_n)\right|^{2}(1/N)\sum_{i=1}^{i=N} (x_{i})^{\kappa}}
{\sum_m \int d\tau_m (x_{i})^{\kappa}\left|\psi_{m}(x_{1},...x_{N},...x_m)\right|^{2}}\,.
\label{em}
\eeq
where $d\tau_n=(dx_i/x_i..dx_N/x_N ...dx_n/x_n)\delta(\sum_{i} x_i-1)$ is the phase volume ;  $n\ge N$ is the 
number of finite-$x$ partons in a given configuration. $R_N$ would be equal to unity  in the case of identical ladders originating from the partons with approximately equal $x_i$. If $N$ is large,  $R_N$  becomes  significantly smaller 
than unity  due to tighter phase volume restrictions in 
the numerator than in the denominator and due to a decrease of the average energy allowed for inelastic collisions.  
  The deviation of $R_N$ from unity   violates AGK combinatorics.

Within the discussed picture, energy-momentum conservation for the final state is realized through a reduction 
of the number of fast spectator constituents in the nucleon with an increase of the number of wounded nucleons 
leading to the strong suppression of production of hadrons in the nucleon fragmentation region and close to the 
central region for large  $N_{coll}$.  In the discussion above, we neglected the contribution of hard interactions into 
the bulk structure of the events. This 
may be an oversimplification for the LHC energies, where the interaction of hard partons with large 
$x_p$ may become black up to the virtualities of few GeV  for central collisions.
This would lead to further suppression of the leading hadron production, $p_t$ broadening of the forward spectrum 
and an additional flow of energy to the central rapidities.

%
%
\end{document}